\documentclass[aps,reprint,showpacs,showkeys,floatfix,longbibliography,superscriptaddress,prstper]{revtex4-1}

\usepackage{graphicx}
\usepackage{dcolumn}
\usepackage{amsmath}
\usepackage{amssymb}
\usepackage{epstopdf}
\usepackage{rotating}
\usepackage{color}
\bibpunct{[}{]}{,}{n}{}{}
\newcommand\Qh{Q_{\textrm{\tiny H}}}
\newcommand\Ql{Q_{\textrm{\tiny L}}}
\newcommand\Th{T_{\textrm{\tiny H}}}
\newcommand\Tl{T_{\textrm{\tiny L}}}
\newcommand\Sh{S_{\textrm{\tiny H}}}
\newcommand\Sl{S_{\textrm{\tiny L}}}
\newcommand\ws{working substance}
\newcommand\etac{\eta_{_{\textrm{\tiny C}}}}
\newcommand\dbar{\mathrm{d}\hspace{-0.5em}^-}
\newcommand\st{$1^{\mathrm{st}}$ }
\newcommand\nd{$2^{\mathrm{nd}}$ }

\newcommand\slt{Second Law of Thermodynamics}

\newcommand\stat{\textit{Stat Mech}}
\newcommand\thermo{\textit{Thermo}}
\newcommand\het{\textit{Heat Engines} tutorial}

\newcommand\eeq{engine entropy question}
\newcommand{\etal}{\textit{et al.}}

\begin{document}

\title{Identifying Student Difficulties with Entropy, Heat Engines, and the Carnot Cycle}


\author{Trevor I. Smith}\affiliation{Department of Physics \& Astronomy and Department of Teacher Education, Rowan University, Glassboro, NJ 08028}

\author{Warren M. Christensen}\affiliation{Department of Physics \& Astronomy, North Dakota State University, Fargo, ND 58102}

\author{Donald B. Mountcastle}\affiliation{Department of Physics \& Astronomy, University of Maine, Orono, ME 04469} 

\author{John R. Thompson}\affiliation{Department of Physics \& Astronomy, University of Maine, Orono, ME 04469}\affiliation{Maine Center for Research in STEM Education, University of Maine, Orono, ME 04469}

\begin{abstract}
We report on several specific student difficulties regarding the Second Law of Thermodynamics in the context of heat engines within upper-division undergraduates thermal physics courses. Data come from ungraded written surveys, graded homework assignments, and videotaped classroom observations of tutorial activities. Written data show that students in these courses do not clearly articulate the connection between the Carnot cycle and the Second Law after lecture instruction. This result is consistent both within and across student populations. Observation data provide evidence for myriad difficulties related to entropy and heat engines, including students' struggles in reasoning about situations that are physically impossible and failures to differentiate between differential and net changes of state properties of a system. Results herein may be seen as the application of previously documented difficulties in the context of heat engines, but others are novel and emphasize the subtle and complex nature of cyclic processes and heat engines, which are central to the teaching and learning of thermodynamics and its applications.  Moreover, the sophistication of these difficulties is indicative of the more advanced thinking required of students at the upper division, whose developing knowledge and understanding give rise to questions and struggles that are inaccessible to novices.
\end{abstract}

\maketitle

\section{Introduction}
\label{sec:intro}
The National Academies recently published a report on the status of discipline-based education research (DBER) that recommends further study of teaching and learning at the upper division as well as topics with interdisciplinary significance \cite{NRCDBER2012}. Because of its relevance in several different domains, thermodynamics is taught across several science and engineering disciplines.  Understanding student ideas about thermodynamics concepts in physics is one important aspect of providing the pedagogical content knowledge needed to teach thermodynamics well, not only in physics, but also in applied disciplines.  Identifying which concepts, representations, and mathematics that students learn well, or with which students struggle, can guide instruction and curriculum development.

In physics, instruction in thermodynamics typically emphasizes an idealized, simplified model: e.g., a closed system of an ideal gas undergoing reversible processes that may include system contact with one or more thermal reservoirs having infinite heat capacities (which allows for heat transfer without change in reservoir temperature). This model system is used to demonstrate the fundamental principles and laws of thermodynamics. Further instruction may explore less ideal systems and processes and compare them to the touchstones of these idealized models.

Two related central topics in thermodynamics are entropy and the \slt.  While quantities related to the First Law of Thermodynamics --- energy, work, and heat transfer --- may appear in previous courses, entropy and the \nd Law most commonly debut in a formal way in a thermodynamics or thermal physics course.  Historically, one of the major applications of entropy and the \nd Law is in the context of heat engines and other thermodynamic cycles: understanding the parameters and the constraints of real engines (such as steam engines) is necessary for proper design and construction. 

The quantity of merit for the performance of a heat engine is known as the thermodynamic efficiency $(\eta)$ and is defined as the work output $(W)$ divided by the heat input $(\Qh)$.  $\eta$ is commonly described as the ratio of ``what you get'' to ``what you pay.''  A critical realization in this area is that the \nd Law places an upper bound on $\eta$ that is not available from application of the \st Law alone. The Carnot cycle or Carnot engine is the thermodynamic cycle that has the maximum efficiency for an engine operating between a given pair of thermal reservoirs, in accordance with the \nd Law.  

As part of an ongoing project, we have examined students' ideas about entropy, heat engines, and the \nd Law. We are particularly interested in the connections students do or do not make between the Carnot cycle, reversibility, and entropy changes as dictated by the \slt. Others have also investigated student ideas regarding entropy \cite{Bucy2006, Christensen2009Student, Bennett2007}, and some have examined this in the context of heat engines \cite{Cochran2006}. In general, findings suggest that many student difficulties persist through introductory and upper-division courses. 

One major goal of our work was the development of a guided-inquiry tutorial to be used in an upper-division thermal physics course to help students connect the Carnot cycle with the limits dictated by the \nd Law. However, this paper focuses on the student difficulties identified as a result of our research. We will present the development and implementation of instructional strategies used to address these difficulties in a separate report.

Expert physicists were all, at one time, advanced undergraduate students; and these advanced undergraduate students were all, at one time, novice introductory students. Due to the in-between status of these advanced undergraduate students, Bing and Redish have described them as ``journeyman physicists'': displaying behaviors similar to novices in certain situations and similar to experts in others \cite{Bing2012}. They find that students in various upper-division undergraduate courses are much more fluid in their choice of problem-solving strategy than introductory students. While introductory students typically only pursue one solution strategy --- even when that strategy becomes unproductive --- upper-division students switch strategies and justifications when their original approach seems to breakdown or reach a dead end. However, many students in the Bing and Redish study failed to come to a satisfactory final solution, indicating that they had not yet mastered the material and cannot be considered expert physicists.

In this paper we present several years' worth of data from written surveys and videotaped classroom observations in advanced undergraduate thermal physics courses that provide evidence for several specific student difficulties. Our written data suggest that many students do not recognize the connection between Carnot's limit on thermodynamic efficiency and the \nd Law. These results are consistent across several years and two institutions. Our observation data strengthen our written data by showing that students do not find this relationship trivial; additionally, we see students struggle with consideration of the physical implications of impossible situations, and differentiation between the differential change of a quantity at an instant and the net change in that quantity over an entire process --- both of which are significant barriers to understanding how the \nd Law applies to heat engines and other complex phenomena. Some of the difficulties we identify are novel; others may be interpreted as the application of previously documented difficulties in the context of heat engines. We begin with an overview of heat engines to introduce our notation and set the stage for the presentation of our data and results.

\section{The Subtleties of Heat Engines}
\label{sec:physics}
A heat engine is a device that converts energy absorbed as heat 
into usable work. To accomplish this, a heat engine requires three things: a high-temperature ($\Th$) thermal reservoir, a low-temperature ($\Tl$) thermal reservoir, and a \ws; e.g., a fixed amount of gas held in a cylinder by a moveable piston). A heat engine operates in a cycle, so that the \ws\ repeatedly returns to its original thermodynamic state. In the course of each cycle, an amount of energy ($\Qh$) is transferred from the $\Th$-reservoir to the \ws; the \ws\ transfers energy to its surroundings by doing work ($W$); and energy ($\Ql$) is transferred from the \ws\ to the $\Tl$-reservoir \footnote{For notation purposes we use the symbols $W$, $\Qh$, and $\Ql$ to represent the absolute values of the energy transfers throughout a heat engine cycle; therefore, they are inherently positive.}.  In the ideal case, both thermal reservoirs have infinite heat capacity and thus maintain constant temperatures.

The application of the First Law of Thermodynamics (\st Law) to each complete cycle yields an expression in terms of these quantities that reflects the energy transfers to and from the working substance: 
\begin{equation}
\Delta U_{\textrm{ws}}=Q_{net}-W_{net}=\Qh-\Ql-W=0,\label{1stlawhe}
\end{equation}
where the final equality results from the \ws\ returning to its original state (defined by equilibrium values of $U$, $V$, $P$, $T$, $S$, etc.). This implies that
\begin{equation}
\Qh-\Ql=W \label{1law2}
\end{equation}
and one may rewrite the thermodynamic efficiency as
\begin{equation}
\eta=\frac{W}{\Qh}=\frac{\Qh-\Ql}{\Qh}=1-\frac{\Ql}{\Qh}.\label{effhe}
\end{equation}

The \nd Law is embodied in the principle of maximizing entropy:
\begin{quote}
The entropy of an isolated system increases in any irreversible [spontaneous, or naturally occurring] process and is unaltered in any reversible [ideal] process \cite[p.\ 96]{Carter2001};
\end{quote}
or mathematically from the entropy inequality,
\begin{equation}
\Delta S_{universe}\geq0,\label{2law1}
\end{equation}
where $S_{universe}$ is the total entropy of the universe, and the equality only holds for ideal reversible processes \footnote{In principle any isolated system could be chosen, but the universe is a natural choice as it is isolated by definition.}. Recognizing that the change in entropy of the working substance will be zero over a complete cycle (because entropy is a state function) and that the temperatures of the reservoirs are constant, the \nd Law in this case takes the form
\begin{equation}
\Delta S_{uni}=\Delta \Sl + \Delta \Sh=\frac{\Ql}{\Tl}-\frac{\Qh}{\Th}\geq0,\label{she}%
\end{equation}
which yields
\begin{equation}
\frac{\Ql}{\Tl}\geq\frac{\Qh}{\Th}.\label{2lawratio}
\end{equation}

Combining Eq.\ \eqref{2lawratio} with the expression for efficiency presented in Eq.\ \eqref{effhe}, one may see that $\eta$ is restricted to the range
\begin{equation}
0<\eta\leq1-\frac{\Tl}{\Th}.\label{eff2}
\end{equation}
The expression on the right side of Eq.\ \eqref{eff2} is defined as Carnot's efficiency ($\etac$) due to the fact that Sadi Carnot proposed a heat engine, consisting of an alternating sequence of (reversible) isothermal and adiabatic processes, that achieves precisely this efficiency. Carnot did not have the benefit of our modern definition of entropy, but his proposed theoretical cycle allows the entropy of the universe to remain unchanged by using only ideal reversible processes.

\section{Background Literature}
\label{sec:lit}

Relatively few studies in PER have focused on students' understanding of topics related to thermal physics at the upper division, but a number of studies have shed light on many student difficulties at the pre-college, introductory undergraduate and advanced undergraduate levels. Many of these studies report students' confusion between the basic concepts of heat and temperature\cite{Harrison1999,Jasien2002,Yeo2001}. 

A handful of studies at the university level have probed students' reasoning and reasoning difficulties with thermodynamic properties from a microscopic perspective, in physics \cite{Kautz2005a,Kautz2005b} and in chemistry \cite{Monteyne2008}.  Overall, these studies point to several inappropriate connections between microscopic and macroscopic properties (e.g., associating particle density with temperature \cite{Kautz2005b}). Other research has shown that students struggle to apply the \st Law correctly in appropriate contexts and suggests student difficulties in recognizing the difference between state variables (e.g., $U$, $S$) and process variables ($W$ and $Q$) \cite{Loverude2002,Meltzer2004,Pollock2007}.

Studies of precollege student ideas about entropy, equilibrium, and reversibility suggest that students entering university physics courses (without having been previously instructed regarding entropy and the \nd Law) have some intuitive ideas about equilibration and irreversibility; however, these ideas are not based on a robust understanding of the concept of entropy as a physicist would define it \cite{Shultz1981,Kesidou1993}. 

Several studies in recent years have focused on student understanding of entropy and the \nd Law in both introductory and upper-division undergraduate physics courses \cite{Christensen2009Student,Thomas1998,Bucy2006,Bucy2007,Bennett2007,Cochran2006}. One prominent finding of this work is students' tendency to use the \nd Law to justify the claim that the entropy of an arbitrary (not necessarily isolated) system must always increase \cite{Christensen2009Student,Thomas1998}. On the other hand, this research has also shown evidence of students treating entropy as a conserved quantity \cite{Christensen2009Student}. In many cases students relate entropy directly (and often incorrectly) to either more familiar thermodynamic quantities (e.g., heat transfer, temperature, work) \cite{Bucy2006,Bucy2007}, or an imprecisely defined sense of ``disorder'' \cite{Bennett2007} when reasoning about entropy changes during particular processes. Langbeheim \etal\ reported similar failures to invoke the \nd Law when asking high school students about the signs of entropy changes during phase separation \cite{Langbeheim2013}. All of these results imply an incomplete understanding of entropy and how to apply the \nd Law.

Cochran \& Heron found that many students (in both introductory and advanced undergraduate courses) did not correctly apply the \nd Law to determine whether or not a proposed heat engine or refrigerator was physically possible \cite{Cochran2006}. Their study focused on students' recognition of the equivalence of various statements of the \nd Law and on developing their abilities to use Carnot's theorem in appropriate contexts.

Overall, while student understanding and application of entropy and the \nd Law have been studied, very little work has looked at the application of these ideas in the context of heat engines, an important touchstone topic in thermodynamics.  Moreover, our current study differs from others in our focus on students' understanding of the physical justification of the mathematical expression for the upper limit on thermodynamic efficiency.

\section{Research Setting and Methods}
\label{sec:methods}
The majority of data for this study were collected in a semester-long upper-division classical thermodynamics course (\thermo) at a 4-year land-grant research university in the northeastern U.S.\ (School 1). The course enrolls approximately 8-12 students each fall semester; the population under investigation was comprised primarily of senior undergraduate physics majors. \thermo\ meets for three 50-minute periods each week. Most instruction is lecture-based, but guided-inquiry tutorials are used in four to seven class periods. Additional data were gathered in a semester-long statistical mechanics course (\stat) offered in the spring semester at School 1 \footnote{Data from the \stat\ course at School 1 were only analyzed for students who had previously taken \thermo.}, and in a semester-long classical and statistical thermal physics course at a 4-year private research university in the northeastern U.S.\ (School 2). Both \thermo\ at School 1 and the statistical thermal physics course at School 2 use Carter's \textit{Classical and Statistical Thermodynamics} as the course textbook \cite{Carter2001}.

Two different written questions were administered, one as an ungraded survey in class (at both schools), and one as a homework problem (at School 1 only).  Students' responses were categorized first by the specific answers given, and second by the explanations provided. Analyzing these explanations, we used a grounded theory approach in which the entire data corpus was examined for common trends, and all data were reexamined to group them into categories defined by these trends \cite{Behrens2004,Strauss1990}. Our objective was to identify and document specific difficulties that students displayed while thinking and reasoning about heat engines. As such, we emphasize the description of students' actions and utterances over our interpretations, and we recognize that any descriptions of students' ideas are our own assumptions based on the data \cite{Heron2003}\label{spdiff}. We often analyzed the data holistically to identify trends across students and data sets.

Video data were collected from classroom observations (at School 1) of students completing guided-inquiry tutorial activities related to heat engines \cite{Smith2009,Smith2011}. These data informed the research in two ways.  First, they provided more depth and complexity to the findings about student ideas and student reasoning from the written data, since students provided more detailed reasoning about physical situations similar to those in the written questions. This serves to strengthen the findings from the written data.  Second, the instructional sequence in the tutorial brought up situations and ideas that weren't covered in the written questions, so additional difficulties were identified.  Data do not exist to verify the prevalence of these difficulties across large populations of students, but we feel their existence is noteworthy. Segments from classroom video episodes were selected for transcription and further analysis based on the content of student discussions. Given our focus on investigating student understanding of particular topics, our methods of gathering video data align with Erickson's \textit{manifest content} approach \cite{Erickson2006}. Each video was watched in its entirety, noting segments that would be interesting and useful for further analysis; these segments were then transcribed along with researcher notes and impressions. Student quotations included in the following sections were selected because they were novel and/or indicative of opinions expressed by the group. Several students made comments and statements that indicated difficulties that were not expected and have not been previously documented. 

More detailed descriptions of the written research instruments are contained in Sec.\ \ref{sec:carnot-rev}, where we present the data collected in each form and interpret the corresponding results.

\section{Connecting the Carnot cycle with the \nd law}
\label{sec:carnot-rev}
As discussed above, Carnot's limit on the thermodynamic efficiency of heat engines can be obtained directly by applying the \st and \nd Laws. One of our major research objectives was to determine whether or not students make this connection and recognize the implications for a heat engine claimed to be operating at an efficiency greater than that of a Carnot engine. 

\begin{figure}[tb]
\begin{center}
\includegraphics[width=3.3in]{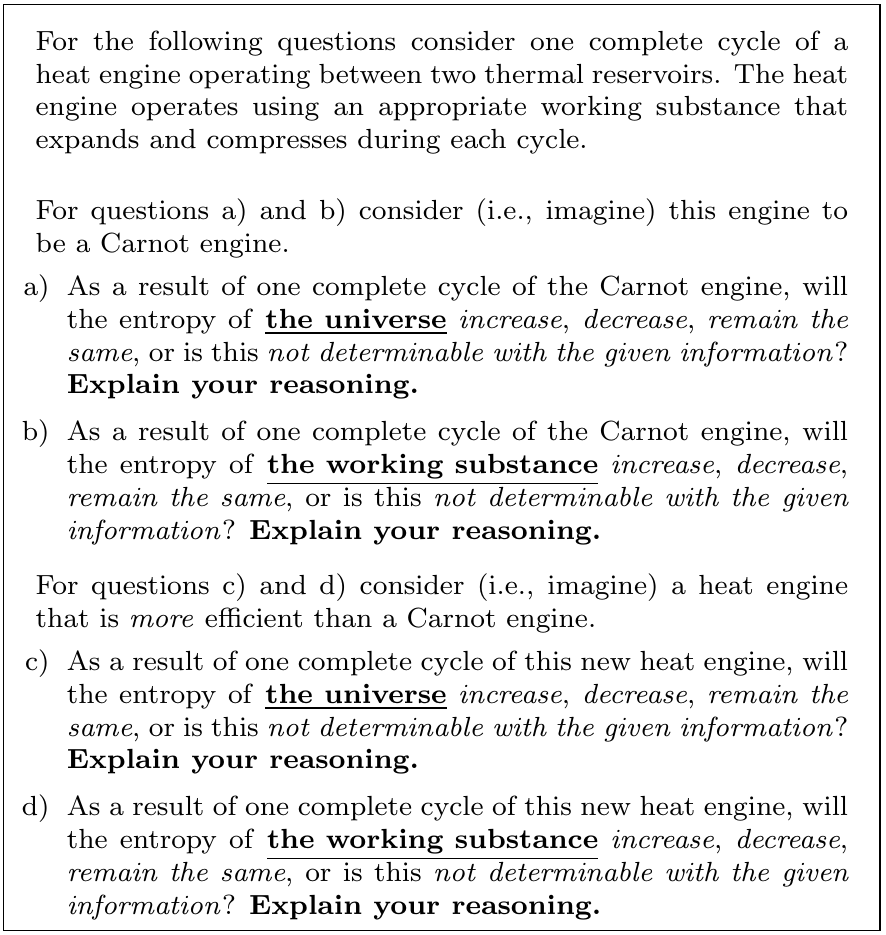}
\end{center}
\caption[The Engine Entropy Question]{The Engine Entropy Question (EEQ). Administered after lecture instruction and again after tutorial instruction.}
\label{eeq}
\end{figure}

The \eeq\ (EEQ, shown in Fig.\ \ref{eeq}) was developed to assess students' understanding of the connection between Carnot's theorem and the \nd Law. The EEQ asks students to consider the change in entropy of various ``systems'' for two heat engines, first as the result of one complete cycle of a Carnot engine, and second as a result of one complete cycle of a heat engine that is hypothetically \textit{more} efficient than the Carnot engine. The students are asked about the change in entropy of the universe (\ws\ and both reservoirs) and then about the change in entropy of the \ws\ alone \footnote{In later versions students were asked about the \ws\ before being asked about the universe.}. 

To fully comprehend the correct answer, the students must understand and apply two ideas: 1) entropy is a state function, and 2) the Carnot cycle is reversible. The fact that entropy is a state function along with the fact that the \ws\ ends the cycle at the same thermodynamic state as it began (by definition of a cycle) indicate that the entropy of the \textit{\ws} must be unchanged after each complete cycle; this statement is true for any ideal heat engine regardless of its efficiency.  The fact that the Carnot cycle is reversible means that the equality must hold in Eq.\ \eqref{2law1}, so the entropy of the \textit{universe} must also remain the same after each complete cycle of a Carnot engine. The fact that the Carnot cycle is reversible also indicates that to obtain a heat engine with an efficiency \textit{greater} than the Carnot efficiency, the \nd Law must be violated. One may conclude that the entropy of the universe would decrease for this better-than-Carnot engine. Thus the correct responses for the EEQ are: (a) same, (b) same, (c) decrease, and (d) same. 

\subsection{Student Responses}
The EEQ was first given to students in \stat, all of whom had previously completed \thermo\ ($N=5$). Several lectures had been spent on heat engines in \thermo, and emphasis was placed on the reversibility of the Carnot cycle. Student responses to the EEQ from this semester indicate that students who had completed a semester-long course on classical thermodynamics did not have a good understanding of the connection between thermodynamic efficiency and changes in entropy: only two students correctly answered all four parts of the EEQ and provided appropriate reasoning for each. 

\begin{figure*}[tb]
\begin{center}
\includegraphics[width=3.4in]{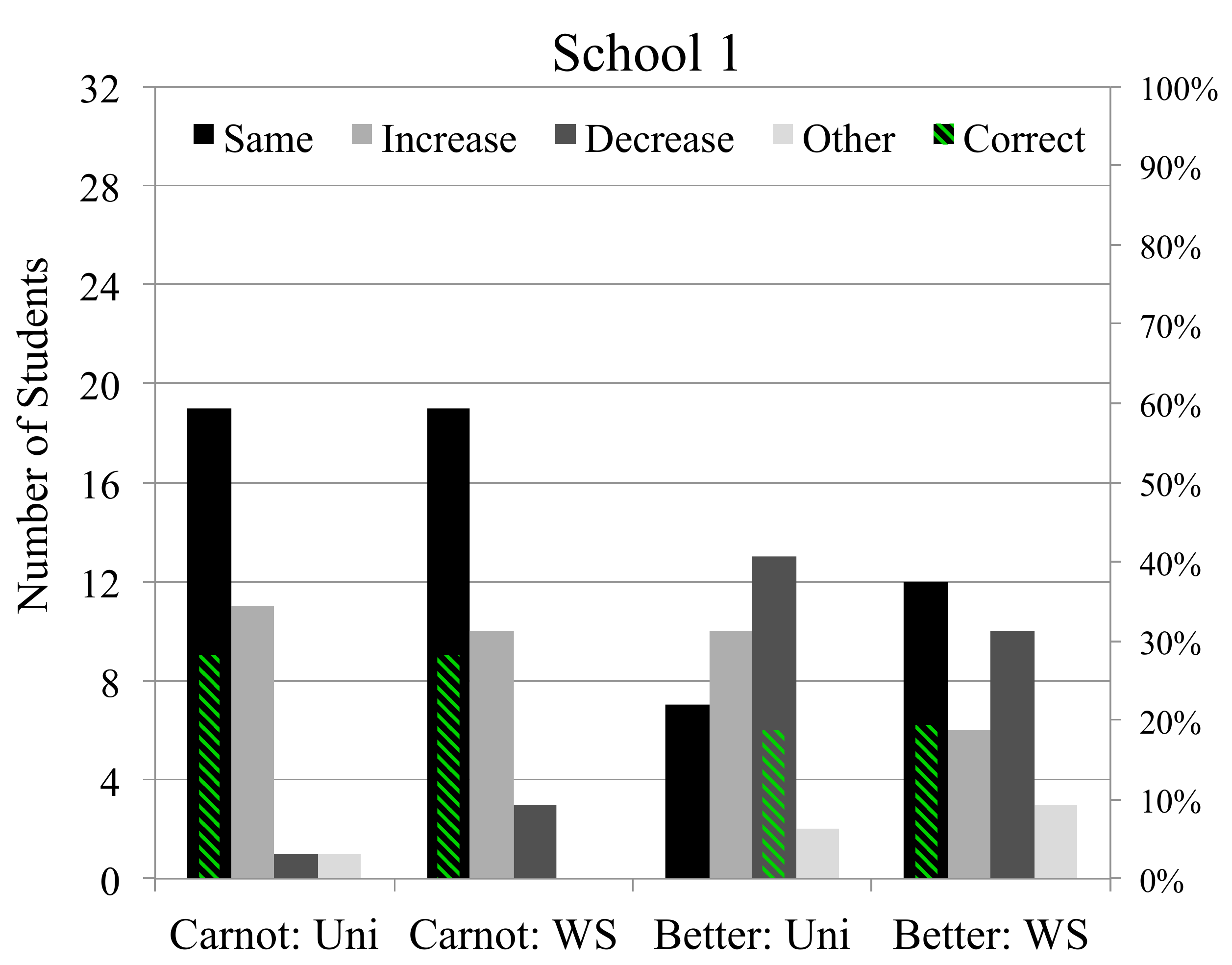}
\hspace{0.15in}
\includegraphics[width=3.4in]{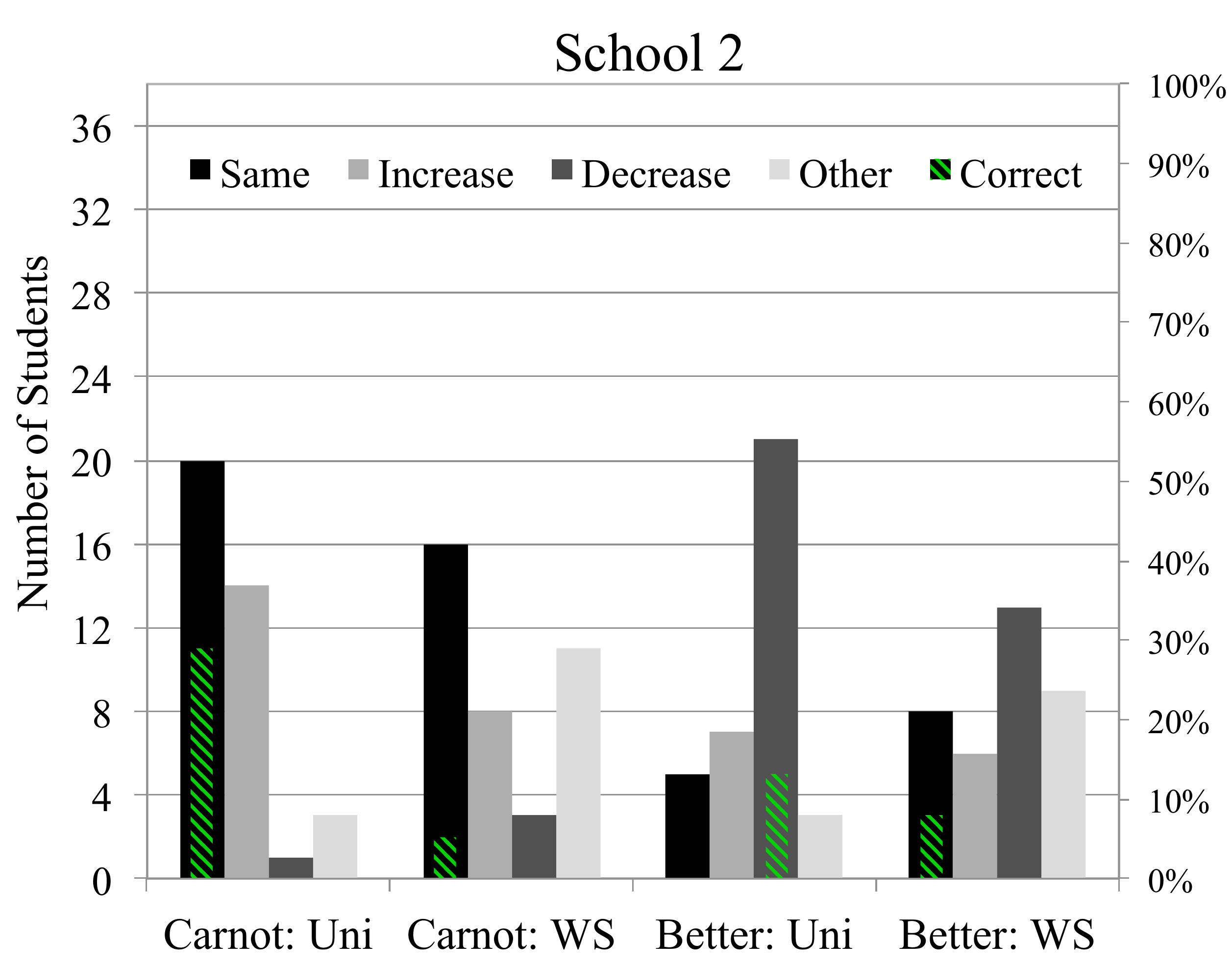}\\
~~~(a)\hspace{3.5in}(b)\\
\includegraphics[width=3.4in]{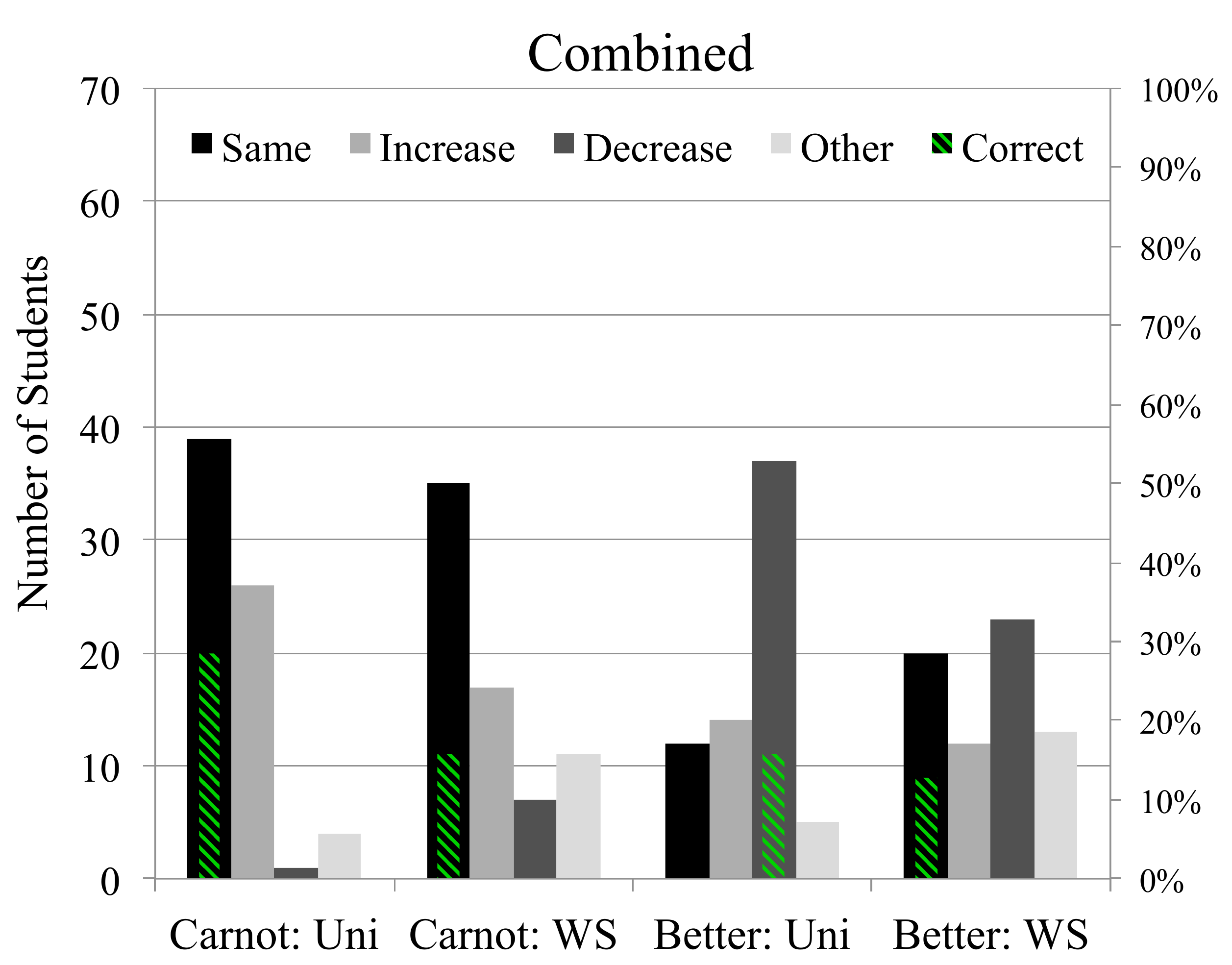}\\
~~(c)
\end{center}
\caption{Response Frequencies on the EEQ Pretest at a) School 1 ($N=32$), b) School 2 ($N=38$), and c) School 1 and School 2 Combined ($N=70$). Green diagonally striped bars indicate students who used correct reasoning. Only four students (two from each school) answered all four parts of the EEQ using correct reasoning. The ``Carnot'' labels refer to questions about the Carnot engine, the ``Better'' labels refer to the questions about an engine that is (supposedly) more efficient than the Carnot engine, the ``Uni'' labels refer to questions regarding the change in entropy of the universe, and the ``WS'' labels refer to questions regarding the change in entropy of the working substance.}
\label{predata}
\end{figure*}

The EEQ was given in \thermo\ for four consecutive years ($N=27$), immediately after all lecture instruction on heat engines. After lecture instruction alone, none of the students used completely correct reasoning for their responses on all four parts of the question. Fig.\ \ref{predata}(a) shows the response frequencies of the aggregate data from all five semesters at School 1. The green diagonally striped bars show the number of students who used correct reasoning for their response on each question.  

The EEQ was also administered once at School 2 approximately three weeks after all (lecture-based) instruction on heat engines ($N=38$). Only two students used completely correct reasoning for their responses on all four parts of the question. Fig.\ \ref{predata}(b) shows the response frequencies from School 2. The data from School 2 appear visually similar to that from School 1, and a Fisher's exact test shows that the two populations are statistically similar in their response patterns for three out of four sub-questions ($p>0.10$, see Table \ref{fetpre}) \cite{Everitt1999,Denenberg1976,Coladarci2004}\footnote{When testing two populations for differences, the threshold for significance was set at $p<0.05$; when testing two populations for similarities, the threshold for significance was set at $p>0.10$. Values of $0.05<p<0.10$ are considered an indication of approaching significance (i.e., not statistically significant but worth mentioning).}. Fig.\ \ref{predata}(c) shows the combined data from both School 1 and School 2 ($N=70$). The most striking result in all plots in Fig.\ \ref{predata} is that only about half of the students who provide the correct answer, justify their choice using correct reasoning (fewer than 30\% of responses include correct reasoning for each sub-question). These data support the hypothesis that many students do not gain a robust understanding of the physical significance of the Carnot cycle and its relationship with the \nd Law after standard instruction.
 
\subsection{Student Reasoning}
\begin{table*}[tb]
\caption[Reasoning on the EEQ Pretest]{Reasoning on the EEQ Pretest. Categories determined by an open analysis of students' written responses to the EEQ pretest.}
\begin{center}
\begin{tabular}{lp{6.5cm}@{~~}p{9cm}}
\hline\hline
Label&Description&Sample student response\\
\hline
{Reversible}&Cite the reversibility of a heat engine&The entropy\dots will stay the same because the process is reversible.\\
{Irreversible}&Cite the irreversibility of a heat engine&As the Carnot cycle is a real process, the entropy of the universe will increase.\\
{Rev.\ + Irr.}&A combination of {Reversible} and {Irreversible} reasoning&I need to know if the processes are reversible. If anything is irreversible then $\Delta S_{uni}>0$.\\
{State Function}& Entropy is a state function; entropy is the same after a complete cycle&Remain the same because $S$ is a state variable and after one cycle the \ws\ is not changed.\\
{Rev.\ + SF}&A combination of {Reversible} and {State Function} reasoning&Entropy will remain the same because it is a complete cycle of a reversible process.\\
{\st Law}&The \st Law is violated and/or energy is not conserved&You get more work out than input.\\
{\nd Law}&The \nd Law is violated&This is contradictory to the \nd Law.\\
{Direction}&The direction in which the device is operated (as a heat engine or a refrigerator) makes a difference&The answer is not determinable because depending on the direction the\dots cycle takes the $\Delta$ entropy could be positive or negative.\\
$\Delta S=Q/T$&Cite that entropy is related to a ratio of heat transfer to temperature&d$S=\dbar\!Q/T$\\
$\Delta S\sim Q$&Entropy is related to heat transfer&Decrease, giving off heat.\\
$\Delta S\sim\Delta T$&Entropy is related to temperature change&The \ws\ is probably going from $\Th$ to $\Tl$ so entropy will be decreasing.\\
\textit{Comparison}&Compare to another heat engine (usually the Carnot engine)&Because a less efficient engine increases entropy, it follows that a more efficient engine decreases entropy.\\
\textit{Statement}&No reasoning given; student merely stated an answer&Decrease.\\
\hline
\end{tabular}
\end{center}
\label{eeq-reason}
\end{table*}%

Shifting focus to examine the reasoning students used when answering the various parts of the EEQ, we have identified ten primary types of reasoning, presented in Table \ref{eeq-reason} \footnote{The ``\textit{Rev.\ + Irr.}'' and ``\textit{Rev.\ + SF}'' categories are considered combinations of types of reasoning and not included in the ten primary types of reasoning. The ``\textit{Statement}'' category is also not counted as a primary type of reasoning because it is evidenced by the absence of reasoning or explanation.}. As described in Sec.\ \ref{sec:methods}, these categories were developed using a grounded theory approach in which we examined the data for common trends and then categorized the data based on these trends. These categories were not suggested by previous research into student understanding of heat engines but derived from the data themselves. Table \ref{eeq-reason} also shows examples of student responses that were categorized as each of the reasoning strategies. These reasoning schemes include considering the (ir)reversibility of a heat engine, the state function property of entropy, and tacitly or explicitly mentioning violations of the \st and/or \nd Laws. 

Some students used more than one of these reasoning types to answer various sub-questions of the EEQ; the \textit{Rev.+Irr.} and \textit{Rev.+SF} categories were created for statistical analyses that indicate combinations of reasoning strategies. Along with those described, one type of response that is closely related to the \textit{Statement} type of reasoning is the statement that the entropy of the universe always increases. This idea was expressed most often (5 out of 64 students) when answering part (a) of the EEQ, and all of these students used the same reasoning or simply stated their answer on part (c). The $\Delta S=Q/T$ reasoning was also accompanied by two related types of reasoning: one case where students related changes in entropy to heat only ($\Delta S\sim Q$), and one in which students relate changes in entropy to changes in temperature ($\Delta S\sim\Delta T$). These reasoning strategies are similar to those seen by Bucy, in which students reason about changes in entropy by discussing either changes in temperature or heat transfer \cite{Bucy2007}. These comparisons may or may not be valid methods for determining entropy change in a particular situation. The reasoning strategies that are considered correct for each sub-question of the EEQ are: a) \textit{Reversible}, b) \textit{State Function}, c) \textit{Violate the \nd Law}, and d) \textit{State Function}. While it is true that the Carnot cycle is reversible and that entropy is a state function, only the former explains why the change in entropy of the universe is zero (part a), while the latter explains why the change in entropy of the \ws\ is zero (part b).

Table \ref{reason-freq} shows the numbers of students at each institution who used each of these lines of reasoning and combinations of reasoning strategies on each sub-question; many categories are only occupied by a handful of students. Moreover, the distribution of the reasoning used differs between School 1 and School 2 on some sub-questions. Using a Fisher's exact test to compare these distributions we found that students at both School 1 and School 2 used similar reasoning on parts (a) and (c). On part (a) this reasoning is most often the correct \textit{Reversible} reasoning, but on part (c) students were most likely to simply state their answer without justifying it in any way (although mentioning violations of the \st and \nd Laws come in a close second, along with \textit{Comparison} reasoning).

\begin{table}[bt]
\caption[Response Frequencies: EEQ Pretest Reasoning]{Response Frequencies: EEQ Pretest Reasoning for School 1 (S1) and School 2 (S2). The ``Uni'' columns refer to questions regarding the change in entropy of the universe, and the ``WS'' columns refer to questions regarding the change in entropy of the working substance. The correct reasoning is shown in bold for each sub-question; the most common reasoning for each population is italicized and shown in blue.}
\begin{center}
\begin{tabular}{l|@{~~}cccccc|cccccc}
\hline\hline
&\multicolumn{5}{c}{Carnot Engine}&&\multicolumn{6}{c}{Better than Carnot}\\
Reasoning&\multicolumn{2}{c}{Uni}&&\multicolumn{2}{c}{WS}&&&\multicolumn{2}{c}{Uni}&&\multicolumn{2}{c}{WS}\\
&S1&S2&&S1&S2&&&S1&S2&&S1&S2\\
\hline
{Reversible}&\textbf{\textit{\color{blue}10}}&\textbf{\textit{\color{blue}12}}&&3&6&&&1&1&&--&1\\
{Irreversible}&2&--&~~~~&--&--&~~&~~&--&--&~~~~&--&--\\
{Rev.\ + Irr.}&1&--&&--&--&&&--&--&&--&--\\
{State Function}&1&--&&\textbf{\textit{\color{blue}6}}&\textbf{2}&&&--&--&&\textbf{\textit{\color{blue}7}}&\textbf{3}\\
{Rev.\ + SF}&1&--&&4&--&&&1&--&&--&--\\
{\st Law}&--&--&&--&--&&&5&1&&2&2\\
{\nd Law}&--&1&&--&--&&&\textbf{5}&\textbf{7}&&--&2\\
{Direction}&--&--&&--&\textit{\color{blue}7}&&&--&--&&--&\textit{\color{blue}8}\\
$\Delta S=Q/T$&3&3&&4&6&&&1&--&&2&1\\
$\Delta S\sim Q$&1&3&&1&\textit{\color{blue}7}&&&1&--&&1&3\\
$\Delta S\sim\Delta T$&--&1&&1&1&&&--&1&&1&--\\
{Comparison}&--&--&&--&--&&&1&8&&1&3\\
{Statement}&1&5&&3&3&&&\textit{\color{blue}6}&\textit{\color{blue}10}&&6&3\\
\hline
\end{tabular}
\end{center}
\label{reason-freq}
\end{table}%

\subsection{Differences between schools}
Although the data are quite similar from both schools, some differences exist. Table \ref{fetpre} shows the results of Fisher's exact tests comparing the distribution of responses at School 1 to those at School 2. The second row shows the results of tests for which all incorrect answers have been combined (including those that gave the ``correct'' answer but did not use correct reasoning). The results of these tests are the same as those comparing the full distribution of responses in that the only significant differences are found when students are asked about the change in entropy of the \ws\ of the Carnot engine.

\begin{table}[tb]
\caption[Fisher's Exact Test: School 1 vs.\ School 2]{Fisher's Exact Test: School 1 vs.\ School 2. Results are $p$-values from Fisher's exact tests comparing the students responses from the two Schools: $p>0.10$ is considered statistically similar. Tests were done on the entire distribution of responses as well as on the distribution if all incorrect responses were combined.}
\begin{center}
\begin{tabular}{cccccc}
\hline\hline
&\multicolumn{2}{c}{Carnot Engine}&~~~&\multicolumn{2}{c}{Better than Carnot}\\
Test&Uni&WS&&Uni&WS\\
\hline
Response&0.95&0.001&&0.78&0.34\\
Correct&1&0.01&&0.34&0.15\\
\hline
\end{tabular}
\end{center}
\label{fetpre}
\end{table}%

On part (b), which asks students about the change in entropy of the \ws\ of a Carnot cycle, the two populations are statistically significantly different ($p=0.001$, see Table \ref{fetpre}). The most salient difference between the two distributions is the large proportion of School 2 students (11 compared to 0 at School 1) who claimed that there is not enough information to answer the question (``other''). In fact, a \textit{post-hoc} Fisher's exact test with the students who answered ``other'' removed yields a result that is statistically similar: $p=0.175$. This shows that the relative distribution of ``increase,'' ``decrease,'' and ``stay the same'' responses is approximately similar and that the difference between the two populations can almost entirely be attributed to some of the School 2 students claiming that not enough information existed to answer the question. A Fisher's exact test also reveals statistically significant differences between the types of reasoning used at each school in parts (b) ($p=0.004$) and (d) ($p=0.02$). Examining Table \ref{reason-freq} one may see that on part (b) students at School 1 most commonly used either the \textit{State Function} (possibly combined with \textit{Reversible}) or the $\Delta S=Q/T$ lines of reasoning, while students at School 2 are most likely to use the $\Delta S=Q/T$, $\Delta S\sim Q$, or \textit{Direction} reasoning. The \textit{Direction} reasoning is particularly interesting as it is quite common at School 2 (for both parts (b) and (d)), but it is not observed at all at School 1. In fact, the same seven students at School 2 used this reasoning on both parts (b) and (d) to say that there was not enough information to determine the change in entropy of the working substance for either engine, indicating consistency across sub-questions, if not correctness. This use of \textit{Direction} reasoning is largely responsible for the comparatively high percentage of students at School 2 claiming that there is not enough information to answer part (b) of the EEQ. This may have been caused in part by the timing of the EEQ at School 2 (three weeks after heat engines instruction). Data do not exist that detail the exact topics discussed during that time, but it is possible that something presented during those classes (e.g., refrigerators) led some students to rely on knowing the direction of a thermodynamic cycle to determine entropy changes. However, we cannot know for certain without more information.

\subsection{Using the Carnot cycle in context: A homework question involving finite reservoirs}
\label{sec:frq}
\begin{figure}[tb]
\begin{center}
\includegraphics[width=3.3in]{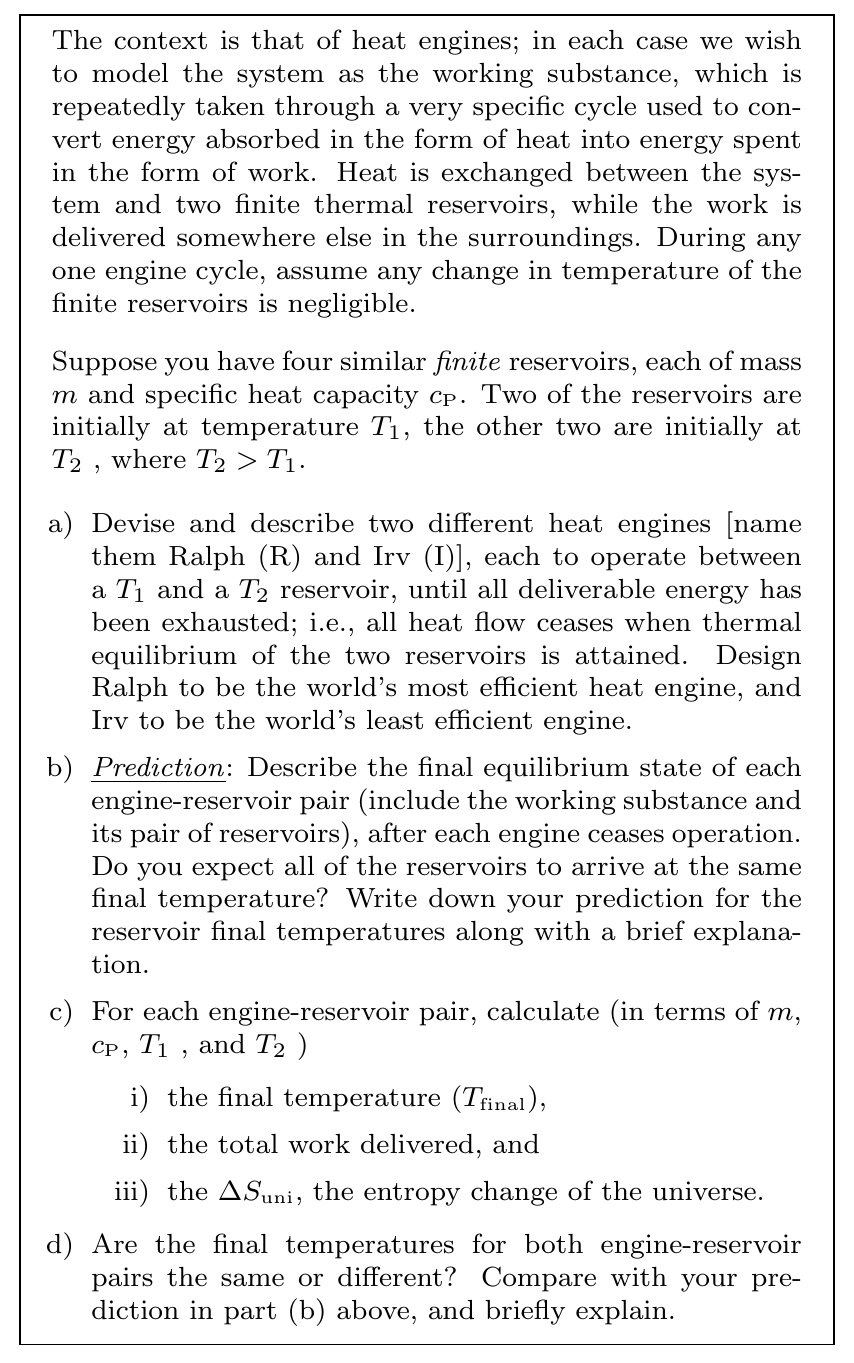}
\end{center}
\caption{The Finite Reservoirs Question (FRQ). Designed based on Carter's problem 7.8 and Baierlein's problem 3.6 \cite{Carter2001,Baierlein1999}. Given as a homework assignment in \thermo\ and in \stat.}
\label{frq}
\end{figure}

In addition to using the EEQ as an ungraded survey, the instructor at School 1 assigned a homework question to specifically assess students' understanding of the Carnot cycle, (ir)reversibility, and the \nd Law. The finite reservoirs question (FRQ) was based on homework problems in texts used in the \thermo\ and \stat\ courses at School 1 (see Fig.\ \ref{frq}).  It was included on a regular homework assignment in four years of \thermo\ ($N=38$), during which time no tutorial instruction was used regarding heat engines.

In the FRQ, students are asked to construct and analyze the most efficient and least efficient heat engines that can operate between pairs of \textit{finite} thermal reservoirs (i.e., with finite specific heat capacity, $c_\textrm{\tiny P}$) with identical initial conditions for the two engines (i.e., the same initial high and low temperatures). 

The solution to part (a) of the FRQ involves students recognizing that the most efficient heat engine (``Ralph'') will have to be a reversible Carnot engine \footnote{The efficiency of the Ralph engine changes with time because the reservoirs have finite heat capacities and changing temperatures. However, in the quasistatic approximation, the temperature of each reservoir does not change appreciably during any one cycle, so Ralph operates at the Carnot efficiency for the current temperatures. After many cycles, the temperatures of the reservoirs will be different, and CarnotÕs efficiency will also be different, but Ralph will always operate at CarnotÕs efficiency, and (more importantly) the combined entropy of the reservoirs remains constant.}
; and similarly that the least efficient heat engine (``Irv'') will do zero work ($\eta=0$). One key element for part (c) is recognizing that Ralph (R) is reversible and thus creates zero entropy ($\Delta S_{uni}=0$). This information may be used to determine both the final temperature of the reservoirs and the total work done by the engine. 

In total, about half of the students for which data exist (20 out of 38) answered the FRQ or a related question correctly after lecture instruction alone. During two years, a preliminary version of the FRQ was used that only looked at the least efficient (Irv) engine ($N=13$; Carter's problem 7.8) \cite[p.\ 124]{Carter2001}. Eight students correctly determined the final temperature of the reservoirs as well as the total change in entropy. 

The full FRQ was used in all other years ($N=25$) \footnote{Some implementations of \stat\ used an expanded version of Baierlein's problem 3.6 that included all parts of the FRQ shown in Fig.\ \ref{frq} \cite[p.\ 72]{Baierlein1999}.}, and only 12 students correctly answered all parts of the question. The most notable result from the remaining students is that even students who realized that Ralph represented the Carnot cycle did not necessarily recognize that the change in entropy of the universe would have to be zero: four students made this error. Another four stated that $\Delta S_{uni}=0$, but did not use it productively to determine the final temperature of the reservoirs. These results are particularly noteworthy, as the uniqueness of the Carnot cycle (and the basis of its importance in thermodynamics) is that it is the \textit{only} reversible heat engine that operates between two thermal reservoirs. 
The prevalence of the disconnect between the Carnot cycle and the \nd Law as well as the failure to apply the specific outcome of the \nd Law to the problem also supports our findings from the EEQ.

\section{Observing Student Difficulties in the Classroom}
\label{sec:observe}
Video data were collected at School 1 (two groups in each of three classes) in order to categorize student reasoning and identify any difficulties that arose during small-group (2-5 students), guided-inquiry activities posed in our \het\ ($N=17$) \cite{Smith2009,Smith2011}. The tutorial presents students with several extreme cases (one in which $\Ql=0$, and one in which $W=0$) before having them consider the limit on thermodynamic efficiency imposed by the first and second laws of thermodynamics. In cases where more than one student displayed a similar difficulty, we have included multiple quotes to allow the reader to evaluate both similarities and differences.

Within our video data we find evidence that students:
\begin{enumerate}
\item have difficulty reasoning about situations that they believe to be impossible,
\item fail to differentiate between differential and net changes of state properties of the working substance,
\item misunderstand the complex differences between state variables and process variables (work and heat transfer), and 
\item neglect the fact that entropy is a state function, instead relating it directly to the more familiar quantities of heat transfer and temperature.
\end{enumerate}
In the following sections we present data to support our claims and also provide counterexamples of student success. 

\subsection{Considering impossible situations}
In an effort to investigate and improve student understanding of heat engines that are (allegedly) more efficient than the Carnot engine (as seen in the \eeq), we asked students to consider a heat engine in which $\Ql=0$. This task is included in our \het; the primary goal of this portion of the tutorial is for students to show explicitly that such an engine would violate the \nd Law (see Eq.\ \eqref{she} with $\Ql=0$) and is, therefore, impossible. Jake (who was working with Gary and Moe) had great difficulty answering these questions, particularly when attempting to reason about changes in entropy due to this cycle \footnote{All student names are pseudonyms.}. Moe proposed the desired response, that the change in entropy of the \ws\ would be zero, that the change in entropy of the reservoirs (and the universe) would be $-\Qh/\Th$, and that this would violate the \nd Law. However, Jake did not agree that the change in entropy of the \ws\ would be zero because he claimed that, in order to convert all of the heat from a single reservoir into work, one cannot use a cyclic process. In this Jake is absolutely correct, which Moe acknowledged by stating (after a very heated discussion in which Moe repeatedly tried to explain his point of view):

\begin{tabular}{r@{~~--~~}p{6.25cm}}
Moe&I'm thinking, \textit{if} it's a cycle, then it can't change all the energy to work. You're thinking, if it's changing all of the heat to work, then it can't be a cycle. We're thinking the same thing for different reasons.\\
Jake&Yeah, alright. We don't know, whatever. Not possible. We don't get this.
\end{tabular}

A similar opinion is observed in Jake's response to parts (c) and (d) of the EEQ pretest, which ask about changes in entropy for a better-than-Carnot heat engine: ``I don't know. We thought the Carnot cycle is the most efficient.'' 

A different group was able to successfully reason about the $\Ql=0$ engine by first considering the fact that the total change in internal energy of the \ws\ over one complete cycle is zero \footnote{For all transcripts, ``I'' stands for the Instructor.}.

\begin{tabular}{r@{~~--~~}p{6.25cm}}
I & So you were trying to relate the change in internal energy to the change in entropy.\\
Dave & Right, which is not going to work.\\
I & But could you say anything\dots\\
Dave & [unintelligible]\\
Sam & Didn't we say the change in entropy for a cycle is zero because it's a state\dots function\dots? On that one over there? [Points to homework assignment]\\
Dave & Yeah, we did.\\
Sam & Yeah, so can't we say for the substance that it goes through a cycle so it has zero change in entropy?\\
Dave & Yeah, that applies for every cycle\dots Yeah.\\
Sam & Yeah.\\
Rick & For the state functions.\\
Sam & [to Rick] Are you buying that?\\
Rick & For state functions or [Sam -- Yeah] for a complete cycle\dots?\\
Sam & Well entropy's a state function so we do  a full cycle on the substance and we're back where we started.\\
Dave & I suppose, but before we've always done one leg of a cycle, but if we're doing it for the whole cycle it would be zero.\\
Sam & Yeah, for the substance, not for\dots[Dave -- Right, yeah] I like that argument.\\
\end{tabular}

\noindent This group went on to correctly determine the change in entropy of the reservoirs (and the universe) to be, $-\Qh/\Th$, and that this violates the \nd Law. In this excerpt the students are able (with instructor support) to fairly quickly apply the state function property of entropy to determine that the change in entropy for the \ws\ is zero (for \textit{all} cycles).

Jake's intuition about situations that can and cannot exist appears to be very strong. In fact, his conviction that the engine for which $\Ql=0$ could not exist is exactly the Kelvin-Planck statement of the \nd Law \cite[p.\ 90]{Carter2001}. Unfortunately, this inability to consider hypothetical and impossible situations may hinder his reasoning abilities in situations in which his intuition is not as well developed. One tool that physicists often use to support a proposition is to show a counter-example that violates known laws of physics. Bing and Redish suggest that the use of imagined (and impossible) situations to gain information about our physical world is a trait of expertise, but that many advanced undergraduate students may not have developed the capacity for this kind of reasoning \cite{Bing2012}. Having students consider the implications of a heat engine that violates the \nd Law encourages this behavior and reasoning skill that is vital for physicists. 

\subsection{Distinguishing between differential change and net change}
In another portion of our \het, students reason about an engine in which $W=0$. Two students (Jonah and Bill) engaged in a particularly interesting conversation when reasoning about the \st Law and efficiency while answering related questions \footnote{All transcriptions of differential vs.\ total change notation (d\underline{\hspace{0.5cm}} vs.\ $\dbar$\underline{\hspace{0.5cm}} vs.\ $\Delta$\underline{\hspace{0.5cm}}) are directly from the video. For example, a transcription of ``$\dbar Q$'' resulted from a verbal utterance of ``dee bar queue.''}: 

\begin{tabular}{r@{~~--~~}p{6.25cm}}
Jonah&What must be true to satisfy the first law, then?\dots [Bill -- uh\dots] well\dots uh\\
Bill&d$Q$ has to be equal to d$U$.\\
Jonah&\textit{Has} to be. d$U$\dots must\dots\\
Bill&so d$Q$ has to be [Jonah -- equal $\dbar Q$] zero.\\
Bill&and d$U$ is zero, so \dots d$Q$ has to be zero\dots\dots That's the only thing I can think of.\\
Jonah&Yeah, I mean, cause d$U$ in a closed cycle, if it's not zero, then you're not conserving energy, so\dots [Bill -- Right.] that's a problem. [Bill -- Yeah.]\\
Bill&It's not a cycle if d$U$ is not zero.\\
Jonah&Yeah.\\
Bill&So, d$Q$ has to be zero\\
Jonah&Yeah.  But thennnn\dots\\
Bill&No. Maybe, maybe it's $\Qh$, or $\Th$ is equal to T-low, $\Tl$.\\
Jonah&Yeah\\
Bill&Because then there'd be no $Q$, [\dots] no heat transfer.\\ 
Jonah&Yeah, but isn't, now isn't the efficiency the work over the heat transfer or something?\\
Bill&Yeah, so it'd be zero.\\
Jonah&Well actually it would kinda be zero over zero wouldn't it? \dots Undefined?\\
Bill&Yeah, I guess.\\
\end{tabular}

During this discussion, Jonah and Bill both refer to cyclic quantities by the differential labels, namely d$U$ instead of $\Delta U$ and $\dbar Q$ instead of $Q_{net}$. Furthermore, Bill uses the incorrect total differential label (d$Q$) while Jonah uses the inexact differential label ($\dbar Q$) for the heat transfer over the entire cycle; nevertheless, they carry on their conversation without being confused by any of these issues.  They have correctly related the heat transfer to the \ws\ and the change in its internal energy by invoking the \st Law ($\dbar Q=\mathrm{d}U$, since $\dbar W=0$), but they have incorrectly determined that there would have to be no heat transfer, requiring the reservoirs to have the same temperature. This has also lead to the enigmatic formulation for efficiency: $\eta=W/\Qh=0/0=\mathrm{undefined}$. 

The error in the conversation had by Bill and Jonah seems to stem from a lack of clarity regarding what they mean by $\dbar Q$ (or d$Q$). It is clear from their exchange that when they use the term ``d$U$,'' they mean the total change in internal energy ($\Delta U$) over a complete cycle: Bill -- ``It's not a cycle if d$U$ is not zero.'' It is also clear that they are using the differential form of the \st Law (d$U=\dbar Q-\dbar W=\dbar Q$) to reason about the situation: Bill -- ``d$Q$ has to be equal to d$U$.'' Combining these pieces of information, Bill and Jonah should have interpreted ``d$Q$'' (and/or ``$\dbar Q$'') to mean the net heat transfer to the working substance over a complete cycle to match their usage of ``d$U$,'' but they did not explicitly make this interpretation. In fact, they interpret d$Q$ as a proxy for any heat transfer when Bill says, ``So, d$Q$ has to be zero,'' and later, ``then there'd be no $Q$, [\dots] no heat transfer.'' In this discussion, Bill and Jonah seem to have no trouble using ``d$U$'' to mean ``$\Delta U$,'' but they do not appropriately apply this definition to interpret ``d$Q$'' as ``$Q_{net}$'' and relating it to the heat transfers to and from the working substance throughout the cycle ($\Qh$ and $\Ql$, respectively). After much discussion the instructor was able to get Bill, Jonah, and Paul (the third group member, who was silent during the above exchange) to realize that they had to consider the net heat transfer as it relates to the various processes that occur throughout the cycle.

The use of imprecise language in terms of differentials and net quantities was not unique to Jonah and Bill. In other groups, Jake stated that ``over a cycle $\mathrm{d}U$ would be zero'' to reason about various proposed definitions of efficiency, and Sam stated that \begin{quote}``\dots d$U$ is zero for the cycle, so $\dbar Q=\dbar W$, which I took it as the net heat is equal to the net work.''\end{quote} Using this reasoning, Sam correctly argued that an alternate definition of efficiency ($\eta=W/[\Qh-\Ql]$) would be unity for all engines. When asked by the instructor to articulate his reasoning again Sam clarified that \begin{quote}``from the \st Law, we know there's no change in energy for the cycle, so d$U$ is zero, so $\dbar Q=\dbar W$, for the whole cycle; so the net heat is equal to the net work.''\end{quote} In this case Sam is incorrectly referring to $\Delta U$ as d$U$ when he states that d$U$=0, but his meaning is clear to his groupmates (as was Jake's): that the total change in internal energy over a cycle is zero. The use of precise language clearly would have benefitted Jonah and Bill, but apparently it was not necessary for Jake or Sam.  

Students' failure to distinguish between differential change (e.g., d$U$) and net change (e.g., $\Delta U$) may be seen as simply a misunderstanding of the calculus. However, many introductory Calculus courses do not present differential quantities as individual entities but rather as part of the notation of the operation of integration. During the introduction of integration in a textbook used for Calculus courses at School 1, the author states, 
\begin{quote}
``The symbol $dx$ has no official meaning by itself; $\int_a^b f(x)dx$ is all one symbol \cite[p.\ 357]{Stewart2001}.''
\end{quote}
This suggests that treating d$U$ as an infinitesimally small version of $\Delta U$ may be a novel task for students in physics classes. This is an example of specific mathematical knowledge that is needed for physics but may not be included in standard mathematics courses. 

Our observation of student difficulties with this relationship aligns with research in mathematics education and in physics education. Findings relevant to this study relate to student responses regarding $dx$ as a quantity. Orton \cite{Orton1983} interviewed calculus students (secondary-level students and preservice mathematics teachers) about differentiation and rates of change. Some students confused $dx$ with the rate of change of $x$, while others described it in a way indistinguishable from a finite increment in $x$. Following an earlier classification of errors in mathematics \cite{Donaldson1963}, these errors were classified primarily as ``structural,'' meaning they were ascribed to difficulties with the relationships between quantities or an essential principle. More recently, Hu and Rebello \cite{Hu2013} identified resources about and conceptual metaphors for differentials used by students in interviews; these include associating the differential with a small amount of a physical quantity or treating it like an object, which is consistent with an earlier report by Artigue \etal\  \cite{Artigue1990}.  Hu and Rebello argue that this object metaphor helps students apply the idea of a differential to physical scenarios, while Artigue \etal\ suggest this approximation approach is used by students ``only as an excuse for loose reasoning'' \cite[p.264]{Artigue1990}. In this light, our observation of students interchanging ``d'' for ``$\Delta$'' in thermodynamics contexts is consistent with a more physical interpretation. 

\subsection{Differentiating between state variables and process variables}
As mentioned in the previous section, Bill incorrectly labeled the differential change in heat d$Q$ (rather than $\dbar Q$). Moreover, Jake was able to interchangeably use ``d$U$'' and ``$\Delta U$'' (as well as ``$\dbar Q$'' and ``$Q$''), but he expressed an insufficient understanding of why $Q$ is not written as $\Delta Q$ given that it is a form of energy transfer, and is opposite $\Delta U$ in the equation of the \st Law. The instructor explained that, notationally, integrating an inexact differential yields a process-dependent quantity (e.g., $\int\dbar Q=Q$), while integrating an exact differential yields a change in a state function (e.g., $\int\mathrm{d}U=\Delta U$); furthermore, the reason that heat has no ``$\Delta$'' symbol is that heat only exists as a process quantity, not as an equilibrium property of a thermodynamic system. This explanation seemed to satisfy Jake, but one may wonder how many other students are disturbed by (or even recognize) this apparent lack of symbolic symmetry and are either unwilling or unable to express their discomfort.


This difficulty is another manifestation of PER findings that many primarily introductory or intermediate-level students reason about work and heat transfer as if they were state functions \cite{Loverude2002,Meltzer2004}. Unlike the prior results in physics, none of our students stated that the work and/or heat transfer over a complete cycle would have to be zero, but Jonah and Sam concluded that the heat transfer must be zero after determining that the net change in internal energy was zero (for a heat engine that does no work). Our students seem to have a better-developed sense of heat and work than the introductory students in other studies, but Jake's confusion about ``$\Delta Q$'' indicates that their ideas may not have coalesced yet into a robust understanding of process variables.

\subsection{Understanding state functions and cycles}
As mentioned above, Sam, Dave, and Rick used the state function property of entropy to correctly determine the net change in entropy over a cycle for a particular heat engine. However, this is not easy for all students: Bonnie and Claude had great difficulty expressing this idea as they worked through our \het. When asked about the change in entropy of the \ws, Claude indicated that $\Delta S_{ws}=(\Qh-\Ql)/T$, but did not have a quick response as to which temperature ``$T$'' represented. After some intervention by the instructor, they agreed that ``$T$'' was the temperature of the \ws\ and that it changed throughout the process (and therefore that their expression couldn't be correct). The instructor proceeded to ask them what it meant for the \ws\ to complete a cycle. Bonnie volunteered that it would return to its original state, and Claude determined that its total change in entropy would have to be zero, because the heat flow would be zero (``$\dbar Q$ would be zero''). Bonnie and Claude only acknowledged the importance of the state function property of entropy and its implications for the working substance after direct instructor intervention. So, even though they eventually ended up at the same point as Sam, Dave, and Rick, their path was much more arduous (and obviously frustrating, as indicated by low voices, sighs, and holding their heads in their hands). Bonnie and Claude's video data and generally poor performance on parts (b) and (d) of the EEQ pretest (which ask about the change in entropy of the \ws\ for each engine) provide evidence that students struggle with the state function property of entropy and how it relates to cyclic processes.

These results are consistent with the findings of Bucy \textit{et al.} that many upper-division students do not correctly use the fact that entropy is a state function to reason about changes, preferring to relate entropy directly to more familiar quantities (like heat transfer and temperature) \cite{Bucy2006,Bucy2007}. This tendency is supported by student use of the $\Delta S=Q/T$ and $\Delta S\sim Q$ lines of reasoning on the EEQ. Many students often rely on mathematical expressions and relationships before thinking about the broader nature of physical quantities.

\section{Conclusions and Implications for Future Work}
We have presented results from an ungraded written survey regarding entropy and efficiency of heat engines (EEQ at Schools 1 and 2), a graded homework assignment regarding heat engines operating between thermal reservoirs with finite heat capacity (FRQ at School 1), and classroom observations of students engaging in guided-inquiry activities posed in our \het\ (School 1). Using multiple data sources has allowed us to deeply explore students' ideas regarding entropy in the context of heat engines and cyclic processes.

The results from both the EEQ and the FRQ show that many students do not demonstrate a robust understanding of the implications of the reversibility of the Carnot cycle with regard to entropy changes after instruction. Only 60\% of respondents to the FRQ (15 out of 25) correctly determined the change in entropy for the universe due to a Carnot cycle operating between two finite reservoirs. Fewer than 30\% of students used correct reasoning while answering the EEQ to determine that the entropy of the universe would stay the same after one complete cycle of a Carnot engine (19 out of 64), and fewer than 20\% of students (11 out of 64) recognized the implication that a heat engine that was more efficient than a Carnot engine would have to violate the laws of thermodynamics and cause the total entropy of the universe to decrease. Moreover, results from Fisher's exact tests show that student responses to the EEQ were remarkably similar at both School 1 and School 2, suggesting difficulties that transcend student population and instructional approach.

These results of students' failure to properly apply the \nd Law in the context of heat engines are consistent with those reported by Cochran \& Heron \cite{Cochran2006}, but our results are unique in that we explicitly asked students to consider the Carnot cycle in the EEQ.

Videorecording classroom episodes of students reasoning about heat engines yielded evidence for several specific difficulties including an inability to reason about situations that they believe to be impossible, misunderstanding the complex and subtle differences between state variables and process variables, and neglecting the state function property of entropy. Many of these may be seen as instantiations of previously documented difficulties \cite{Christensen2009Student, Meltzer2004, Bucy2006, Bucy2007, Hu2013, Loverude2002, Orton1983, Artigue1990, Donaldson1963, Bing2012}; however, their application in the context of heat engines provides evidence for the widespread nature of these difficulties and uncovers previously undocumented difficulties. 

Of particular interest is the evidence on students' failure to differentiate between differential and net changes. Using differentials to represent infinitesimal changes of physical quantities deviates from the typical practice in many calculus classes; further study is warranted to determine students' understanding of differentials in physical contexts.

The two aspects of the research presented here complement and reinforce each other. Written survey results show the prevalence and consistency of several student difficulties related to entropy across two different student populations. The FRQ homework data show that these difficulties manifest in different contexts within the same population and provide further evidence of their tenacity. On the other hand, the observation data provide evidence for the existence of myriad other student difficulties; many of these were expressed by only a handful of students, but none was expressed by only a single student, and all episodes required intervention either by the instructor or other students for those who expressed the difficulty to move past it. This suggests that these difficulties are robust and may, at different times, be expressed by a significant portion of the student population. Further investigation is needed to determine how prevalent these difficulties are within the broader population of upper-division physics students.

Our work provides significant insight into students' understanding of heat engines, especially at the upper division. 
These results provide additional evidence for the subtle and complex nature of heat engines and cyclic processes, which are central to the teaching and learning of thermodynamics and its applications. Moreover, the sophistication of students' difficulties is indicative of ``journeyman physicists,'' whose developing knowledge and understanding give rise to questions and struggles that are inaccessible to novices \cite{Bing2012}.

Part of our work involves the development of a guided-inquiry tutorial to help students better understand heat engines and the connection between Carnot's theorem and the \nd Law.  As mentioned, the video data were collected during implementation of the tutorial in class, allowing us to have another method for investigating student reasoning about these topics while addressing specific difficulties we have identified in both the written and video data.  We have discussed preliminary results demonstrating the positive impact our \het\ has on students' ideas related to heat engines and entropy \cite{Smith2009}; however, more work is needed here as well to broaden the research population and investigate our tutorial's effectiveness in multiple classroom settings.

\begin{acknowledgments}
We thank members of the University of Maine Physics Education Research Laboratory, and our colleagues Michael Loverude and David Meltzer for their continued collaboration and feedback on this work. We are deeply indebted to the instructors of the thermal physics courses in which data were collected. We also thank three anonymous referees for their constructive feedback. This material is based upon work partially supported by the National Science Foundation under Grant No.\ DUE-0817282 and by the Maine Academic Prominence Initiative, for which we are grateful. 
\end{acknowledgments}

\bibliography{TIS}

\begin{thebibliography}{47}%
\makeatletter
\providecommand \@ifxundefined [1]{%
 \@ifx{#1\undefined}
}%
\providecommand \@ifnum [1]{%
 \ifnum #1\expandafter \@firstoftwo
 \else \expandafter \@secondoftwo
 \fi
}%
\providecommand \@ifx [1]{%
 \ifx #1\expandafter \@firstoftwo
 \else \expandafter \@secondoftwo
 \fi
}%
\providecommand \natexlab [1]{#1}%
\providecommand \enquote  [1]{``#1''}%
\providecommand \bibnamefont  [1]{#1}%
\providecommand \bibfnamefont [1]{#1}%
\providecommand \citenamefont [1]{#1}%
\providecommand \href@noop [0]{\@secondoftwo}%
\providecommand \href [0]{\begingroup \@sanitize@url \@href}%
\providecommand \@href[1]{\@@startlink{#1}\@@href}%
\providecommand \@@href[1]{\endgroup#1\@@endlink}%
\providecommand \@sanitize@url [0]{\catcode `\\12\catcode `\$12\catcode
  `\&12\catcode `\#12\catcode `\^12\catcode `\_12\catcode `\%12\relax}%
\providecommand \@@startlink[1]{}%
\providecommand \@@endlink[0]{}%
\providecommand \url  [0]{\begingroup\@sanitize@url \@url }%
\providecommand \@url [1]{\endgroup\@href {#1}{\urlprefix }}%
\providecommand \urlprefix  [0]{URL }%
\providecommand \Eprint [0]{\href }%
\providecommand \doibase [0]{http://dx.doi.org/}%
\providecommand \selectlanguage [0]{\@gobble}%
\providecommand \bibinfo  [0]{\@secondoftwo}%
\providecommand \bibfield  [0]{\@secondoftwo}%
\providecommand \translation [1]{[#1]}%
\providecommand \BibitemOpen [0]{}%
\providecommand \bibitemStop [0]{}%
\providecommand \bibitemNoStop [0]{.\EOS\space}%
\providecommand \EOS [0]{\spacefactor3000\relax}%
\providecommand \BibitemShut  [1]{\csname bibitem#1\endcsname}%
\let\auto@bib@innerbib\@empty
\bibitem [{\citenamefont {Singer}\ \emph {et~al.}(2012)\citenamefont {Singer},
  \citenamefont {Nielsen},\ and\ \citenamefont {Schweingruber}}]{NRCDBER2012}%
  \BibitemOpen
  \bibinfo {editor} {\bibfnamefont {Susan~R.}\ \bibnamefont {Singer}}, \bibinfo
  {editor} {\bibfnamefont {Natalie~R.}\ \bibnamefont {Nielsen}}, \ and\
  \bibinfo {editor} {\bibfnamefont {Heidi~A.}\ \bibnamefont {Schweingruber}},\
  eds.,\ \href@noop {} {\emph {\bibinfo {title} {Discipline-Based Education
  Research: Understanding and Improving Learning in Undergraduate Science and
  Engineering}}}\ (\bibinfo  {publisher} {The National Academies Press},\
  \bibinfo {address} {Washington, D.C.},\ \bibinfo {year} {2012})\BibitemShut
  {NoStop}%
\bibitem [{\citenamefont {Bucy}\ \emph {et~al.}(2006)\citenamefont {Bucy},
  \citenamefont {Thompson},\ and\ \citenamefont {Mountcastle}}]{Bucy2006}%
  \BibitemOpen
  \bibfield  {author} {\bibinfo {author} {\bibfnamefont {Brandon~R.}\
  \bibnamefont {Bucy}}, \bibinfo {author} {\bibfnamefont {John~R.}\
  \bibnamefont {Thompson}}, \ and\ \bibinfo {author} {\bibfnamefont
  {Donald~B.}\ \bibnamefont {Mountcastle}},\ }\bibfield  {title} {\enquote
  {\bibinfo {title} {What is entropy? {A}dvanced undergraduate performance
  comparing ideal gas processes},}\ }in\ \href {\doibase 10.1063/1.2177028}
  {\emph {\bibinfo {booktitle} {2005 Physics Education Research Conference, AIP
  Conference Proceedings}}},\ Vol.\ \bibinfo {volume} {818},\ \bibinfo {editor}
  {edited by\ \bibinfo {editor} {\bibfnamefont {Paula}\ \bibnamefont {Heron}},
  \bibinfo {editor} {\bibfnamefont {Laura}\ \bibnamefont {McCullough}}, \ and\
  \bibinfo {editor} {\bibfnamefont {Jeffrey}\ \bibnamefont {Marx}}},\ \bibinfo
  {organization} {American Association of Physics Teachers}\ (\bibinfo
  {publisher} {American Institute of Physics},\ \bibinfo {address} {Melville,
  NY},\ \bibinfo {year} {2006})\ pp.\ \bibinfo {pages} {81--84}\BibitemShut
  {NoStop}%
\bibitem [{\citenamefont {Christensen}\ \emph {et~al.}(2009)\citenamefont
  {Christensen}, \citenamefont {Meltzer},\ and\ \citenamefont
  {Ogilvie}}]{Christensen2009Student}%
  \BibitemOpen
  \bibfield  {author} {\bibinfo {author} {\bibfnamefont {Warren~M.}\
  \bibnamefont {Christensen}}, \bibinfo {author} {\bibfnamefont {David~E.}\
  \bibnamefont {Meltzer}}, \ and\ \bibinfo {author} {\bibfnamefont {C.~A.}\
  \bibnamefont {Ogilvie}},\ }\bibfield  {title} {\enquote {\bibinfo {title}
  {Student ideas regarding entropy and the second law of thermodynamics in an
  introductory physics course},}\ }\href
  {http://scitation.aip.org/getabs/servlet/GetabsServlet?prog=normal\&id=AJPIAS000077000010000907000001\&idtype=cvips\&gifs=yes}
  {\bibfield  {journal} {\bibinfo  {journal} {American Journal of Physics}\
  }\textbf {\bibinfo {volume} {77}},\ \bibinfo {pages} {907--917} (\bibinfo
  {year} {2009})}\BibitemShut {NoStop}%
\bibitem [{\citenamefont {Bennett}\ and\ \citenamefont
  {S\"{o}zbilir}(2007)}]{Bennett2007}%
  \BibitemOpen
  \bibfield  {author} {\bibinfo {author} {\bibfnamefont {Judith~M.}\
  \bibnamefont {Bennett}}\ and\ \bibinfo {author} {\bibfnamefont {Mustafa}\
  \bibnamefont {S\"{o}zbilir}},\ }\bibfield  {title} {\enquote {\bibinfo
  {title} {A study of turkish chemistry undergraduates' understanding of
  entropy},}\ }\href@noop {} {\bibfield  {journal} {\bibinfo  {journal}
  {Journal of Chemical Education}\ }\textbf {\bibinfo {volume} {84}},\ \bibinfo
  {pages} {1204--1208} (\bibinfo {year} {2007})}\BibitemShut {NoStop}%
\bibitem [{\citenamefont {Cochran}\ and\ \citenamefont
  {Heron}(2006)}]{Cochran2006}%
  \BibitemOpen
  \bibfield  {author} {\bibinfo {author} {\bibfnamefont {Matthew~J.}\
  \bibnamefont {Cochran}}\ and\ \bibinfo {author} {\bibfnamefont {Paula R.~L.}\
  \bibnamefont {Heron}},\ }\bibfield  {title} {\enquote {\bibinfo {title}
  {Development and assessment of research-based tutorials on heat engines and
  the second law of thermodynamics},}\ }\href@noop {} {\bibfield  {journal}
  {\bibinfo  {journal} {Am. J. Phys.}\ }\textbf {\bibinfo {volume} {74}},\
  \bibinfo {pages} {734--741} (\bibinfo {year} {2006})}\BibitemShut {NoStop}%
\bibitem [{\citenamefont {Bing}\ and\ \citenamefont {Redish}(2012)}]{Bing2012}%
  \BibitemOpen
  \bibfield  {author} {\bibinfo {author} {\bibfnamefont {Thomas~J.}\
  \bibnamefont {Bing}}\ and\ \bibinfo {author} {\bibfnamefont {Edward~F.}\
  \bibnamefont {Redish}},\ }\bibfield  {title} {\enquote {\bibinfo {title}
  {Epistemic complexity and the journeyman-expert transition},}\ }\href@noop {}
  {\bibfield  {journal} {\bibinfo  {journal} {Phys. Rev. ST Phys. Educ. Res.}\
  }\textbf {\bibinfo {volume} {8}},\ \bibinfo {pages} {010105} (\bibinfo {year}
  {2012})}\BibitemShut {NoStop}%
\bibitem [{Note1()}]{Note1}%
  \BibitemOpen
  \bibinfo {note} {For notation purposes we use the symbols $W$, $Q_{\protect
  \textrm {\relax \protect \fontsize {5}{6}\protect \selectfont H}}$, and
  $Q_{\protect \textrm {\relax \protect \fontsize {5}{6}\protect \selectfont
  L}}$ to represent the absolute values of the energy transfers throughout a
  heat engine cycle; therefore, they are inherently positive.}\BibitemShut
  {Stop}%
\bibitem [{\citenamefont {Carter}(2001)}]{Carter2001}%
  \BibitemOpen
  \bibfield  {author} {\bibinfo {author} {\bibfnamefont {Ashley~H.}\
  \bibnamefont {Carter}},\ }\href@noop {} {\emph {\bibinfo {title} {Classical
  and Statistical Thermodynamics}}}\ (\bibinfo  {publisher} {Prentice Hall},\
  \bibinfo {address} {Upper Saddle River, NJ},\ \bibinfo {year}
  {2001})\BibitemShut {NoStop}%
\bibitem [{Note2()}]{Note2}%
  \BibitemOpen
  \bibinfo {note} {In principle any isolated system could be chosen, but the
  universe is a natural choice as it is isolated by definition.}\BibitemShut
  {Stop}%
\bibitem [{\citenamefont {Harrison}\ \emph {et~al.}(1999)\citenamefont
  {Harrison}, \citenamefont {Grayson},\ and\ \citenamefont
  {Treagust}}]{Harrison1999}%
  \BibitemOpen
  \bibfield  {author} {\bibinfo {author} {\bibfnamefont {Allan~G.}\
  \bibnamefont {Harrison}}, \bibinfo {author} {\bibfnamefont {Diane~J.}\
  \bibnamefont {Grayson}}, \ and\ \bibinfo {author} {\bibfnamefont {David~F.}\
  \bibnamefont {Treagust}},\ }\bibfield  {title} {\enquote {\bibinfo {title}
  {Investigating a grade 11 student's evolving conceptions of heat and
  temperature},}\ }\href@noop {} {\bibfield  {journal} {\bibinfo  {journal}
  {Journal of Research in Science Teaching}\ }\textbf {\bibinfo {volume}
  {36}},\ \bibinfo {pages} {55--87} (\bibinfo {year} {1999})}\BibitemShut
  {NoStop}%
\bibitem [{\citenamefont {Jasien}\ and\ \citenamefont
  {Oberem}(2002)}]{Jasien2002}%
  \BibitemOpen
  \bibfield  {author} {\bibinfo {author} {\bibfnamefont {Paul~G.}\ \bibnamefont
  {Jasien}}\ and\ \bibinfo {author} {\bibfnamefont {Graham~E.}\ \bibnamefont
  {Oberem}},\ }\bibfield  {title} {\enquote {\bibinfo {title} {Understanding of
  elementary concepts in heat and temperature among college students and
  {K--12} teachers},}\ }\href@noop {} {\bibfield  {journal} {\bibinfo
  {journal} {Journal of Chemical Education}\ }\textbf {\bibinfo {volume}
  {79}},\ \bibinfo {pages} {889--895} (\bibinfo {year} {2002})}\BibitemShut
  {NoStop}%
\bibitem [{\citenamefont {Yeo}\ and\ \citenamefont {Zadnik}(2001)}]{Yeo2001}%
  \BibitemOpen
  \bibfield  {author} {\bibinfo {author} {\bibfnamefont {Shelley}\ \bibnamefont
  {Yeo}}\ and\ \bibinfo {author} {\bibfnamefont {Marjan}\ \bibnamefont
  {Zadnik}},\ }\bibfield  {title} {\enquote {\bibinfo {title} {Introductory
  thermal concept evaluation: Assessing students' understanding},}\ }\href@noop
  {} {\bibfield  {journal} {\bibinfo  {journal} {The Physics Teacher}\ }\textbf
  {\bibinfo {volume} {39}},\ \bibinfo {pages} {496--504} (\bibinfo {year}
  {2001})}\BibitemShut {NoStop}%
\bibitem [{\citenamefont {Kautz}\ \emph
  {et~al.}(2005{\natexlab{a}})\citenamefont {Kautz}, \citenamefont {Heron},
  \citenamefont {Loverude},\ and\ \citenamefont {McDermott}}]{Kautz2005a}%
  \BibitemOpen
  \bibfield  {author} {\bibinfo {author} {\bibfnamefont {Chrstian~H.}\
  \bibnamefont {Kautz}}, \bibinfo {author} {\bibfnamefont {Paula R.~L.}\
  \bibnamefont {Heron}}, \bibinfo {author} {\bibfnamefont {Michael~E.}\
  \bibnamefont {Loverude}}, \ and\ \bibinfo {author} {\bibfnamefont
  {Lillian~C.}\ \bibnamefont {McDermott}},\ }\bibfield  {title} {\enquote
  {\bibinfo {title} {Student understanding of the ideal gas law, part i: A
  macroscopic perspective},}\ }\href@noop {} {\bibfield  {journal} {\bibinfo
  {journal} {American Journal of Physics}\ }\textbf {\bibinfo {volume} {73}},\
  \bibinfo {pages} {1055--1063} (\bibinfo {year}
  {2005}{\natexlab{a}})}\BibitemShut {NoStop}%
\bibitem [{\citenamefont {Kautz}\ \emph
  {et~al.}(2005{\natexlab{b}})\citenamefont {Kautz}, \citenamefont {Heron},
  \citenamefont {Shaffer},\ and\ \citenamefont {McDermott}}]{Kautz2005b}%
  \BibitemOpen
  \bibfield  {author} {\bibinfo {author} {\bibfnamefont {Chrstian~H.}\
  \bibnamefont {Kautz}}, \bibinfo {author} {\bibfnamefont {Paula R.~L.}\
  \bibnamefont {Heron}}, \bibinfo {author} {\bibfnamefont {Peter~S.}\
  \bibnamefont {Shaffer}}, \ and\ \bibinfo {author} {\bibfnamefont
  {Lillian~C.}\ \bibnamefont {McDermott}},\ }\bibfield  {title} {\enquote
  {\bibinfo {title} {Student understanding of the ideal gas law, part ii: A
  microscopic perspective},}\ }\href@noop {} {\bibfield  {journal} {\bibinfo
  {journal} {American Journal of Physics}\ }\textbf {\bibinfo {volume} {73}},\
  \bibinfo {pages} {1064--1071} (\bibinfo {year}
  {2005}{\natexlab{b}})}\BibitemShut {NoStop}%
\bibitem [{\citenamefont {Monteyne}\ \emph {et~al.}(2008)\citenamefont
  {Monteyne}, \citenamefont {Gonzalez},\ and\ \citenamefont
  {Loverude}}]{Monteyne2008}%
  \BibitemOpen
  \bibfield  {author} {\bibinfo {author} {\bibfnamefont {Kereen}\ \bibnamefont
  {Monteyne}}, \bibinfo {author} {\bibfnamefont {Barbara~L.}\ \bibnamefont
  {Gonzalez}}, \ and\ \bibinfo {author} {\bibfnamefont {Michael~E.}\
  \bibnamefont {Loverude}},\ }\bibfield  {title} {\enquote {\bibinfo {title}
  {An interdiscplinary study of student ability to connect particulate and
  macroscopic representations of a gas},}\ }in\ \href@noop {} {\emph {\bibinfo
  {booktitle} {2008 Physics Education Research Conference, AIP Conference
  Proceedings}}},\ Vol.\ \bibinfo {volume} {818},\ \bibinfo {editor} {edited
  by\ \bibinfo {editor} {\bibfnamefont {Charles}\ \bibnamefont {Henderson}},
  \bibinfo {editor} {\bibfnamefont {Mel}\ \bibnamefont {Sabella}}, \ and\
  \bibinfo {editor} {\bibfnamefont {Leon}\ \bibnamefont {Hsu}}},\ \bibinfo
  {organization} {American Association of Physics Teachers}\ (\bibinfo
  {publisher} {American Institute of Physics},\ \bibinfo {address} {Melville,
  NY},\ \bibinfo {year} {2008})\ pp.\ \bibinfo {pages} {163--166}\BibitemShut
  {NoStop}%
\bibitem [{\citenamefont {Loverude}\ \emph {et~al.}(2002)\citenamefont
  {Loverude}, \citenamefont {Kautz},\ and\ \citenamefont
  {Heron}}]{Loverude2002}%
  \BibitemOpen
  \bibfield  {author} {\bibinfo {author} {\bibfnamefont {Michael~E.}\
  \bibnamefont {Loverude}}, \bibinfo {author} {\bibfnamefont {Chrstian~H.}\
  \bibnamefont {Kautz}}, \ and\ \bibinfo {author} {\bibfnamefont {Paula R.~L.}\
  \bibnamefont {Heron}},\ }\bibfield  {title} {\enquote {\bibinfo {title}
  {Student understanding of the first law of thermodynamics: Relating work to
  the adiabatic compression of an ideal gas},}\ }\href@noop {} {\bibfield
  {journal} {\bibinfo  {journal} {American Journal of Physics}\ }\textbf
  {\bibinfo {volume} {70}},\ \bibinfo {pages} {137--148} (\bibinfo {year}
  {2002})}\BibitemShut {NoStop}%
\bibitem [{\citenamefont {Meltzer}(2004)}]{Meltzer2004}%
  \BibitemOpen
  \bibfield  {author} {\bibinfo {author} {\bibfnamefont {David~E.}\
  \bibnamefont {Meltzer}},\ }\bibfield  {title} {\enquote {\bibinfo {title}
  {Investigation of students' reasoning regarding heat, work, and the first law
  of thermodynamics in an introductory calculus-based general physics
  course},}\ }\href@noop {} {\bibfield  {journal} {\bibinfo  {journal}
  {American Journal of Physics}\ }\textbf {\bibinfo {volume} {72}},\ \bibinfo
  {pages} {1432--1446} (\bibinfo {year} {2004})}\BibitemShut {NoStop}%
\bibitem [{\citenamefont {Pollock}\ \emph {et~al.}(2007)\citenamefont
  {Pollock}, \citenamefont {Thompson},\ and\ \citenamefont
  {Mountcastle}}]{Pollock2007}%
  \BibitemOpen
  \bibfield  {author} {\bibinfo {author} {\bibfnamefont {Evan~B.}\ \bibnamefont
  {Pollock}}, \bibinfo {author} {\bibfnamefont {John~R.}\ \bibnamefont
  {Thompson}}, \ and\ \bibinfo {author} {\bibfnamefont {Donald~B.}\
  \bibnamefont {Mountcastle}},\ }\bibfield  {title} {\enquote {\bibinfo {title}
  {Student understanding of the physics and mathematics of process variables in
  p-v diagrams},}\ }in\ \href {\doibase 10.1063/1.2820924} {\emph {\bibinfo
  {booktitle} {2007 Physics Education Research Conference, AIP Conference
  Proceedings}}},\ Vol.\ \bibinfo {volume} {951},\ \bibinfo {editor} {edited
  by\ \bibinfo {editor} {\bibfnamefont {Leon}\ \bibnamefont {Hsu}}, \bibinfo
  {editor} {\bibfnamefont {Charles}\ \bibnamefont {Henderson}}, \ and\ \bibinfo
  {editor} {\bibfnamefont {Laura}\ \bibnamefont {McCullough}}},\ \bibinfo
  {organization} {American Association of Physics Teachers}\ (\bibinfo
  {publisher} {American Institute of Physics},\ \bibinfo {address} {Melville,
  NY},\ \bibinfo {year} {2007})\ pp.\ \bibinfo {pages} {168--171}\BibitemShut
  {NoStop}%
\bibitem [{\citenamefont {Schultz}\ and\ \citenamefont
  {Coddington}(1981)}]{Shultz1981}%
  \BibitemOpen
  \bibfield  {author} {\bibinfo {author} {\bibfnamefont {Thomas~R.}\
  \bibnamefont {Schultz}}\ and\ \bibinfo {author} {\bibfnamefont {Marilyn}\
  \bibnamefont {Coddington}},\ }\bibfield  {title} {\enquote {\bibinfo {title}
  {Development of the concepts of energy conservation and entropy},}\
  }\href@noop {} {\bibfield  {journal} {\bibinfo  {journal} {Journal of
  Experimental Child Psychology}\ }\textbf {\bibinfo {volume} {31}},\ \bibinfo
  {pages} {131--153} (\bibinfo {year} {1981})}\BibitemShut {NoStop}%
\bibitem [{\citenamefont {Kesidou}\ and\ \citenamefont
  {Duit}(1993)}]{Kesidou1993}%
  \BibitemOpen
  \bibfield  {author} {\bibinfo {author} {\bibfnamefont {Sofia}\ \bibnamefont
  {Kesidou}}\ and\ \bibinfo {author} {\bibfnamefont {Reinders}\ \bibnamefont
  {Duit}},\ }\bibfield  {title} {\enquote {\bibinfo {title} {Students'
  conceptions of the second law of thermodynamics --- an interprative study},}\
  }\href@noop {} {\bibfield  {journal} {\bibinfo  {journal} {Journal of
  Research in Science Teaching}\ }\textbf {\bibinfo {volume} {30}},\ \bibinfo
  {pages} {85--106} (\bibinfo {year} {1993})}\BibitemShut {NoStop}%
\bibitem [{\citenamefont {Thomas}\ and\ \citenamefont
  {Schwenz}(1998)}]{Thomas1998}%
  \BibitemOpen
  \bibfield  {author} {\bibinfo {author} {\bibfnamefont {Peter~L.}\
  \bibnamefont {Thomas}}\ and\ \bibinfo {author} {\bibfnamefont {Richard~W.}\
  \bibnamefont {Schwenz}},\ }\bibfield  {title} {\enquote {\bibinfo {title}
  {College physical chemistry students' conceptions of equilibrium and
  fundamental thermodynamics},}\ }\href@noop {} {\bibfield  {journal} {\bibinfo
   {journal} {Journal of Research in Science Teaching}\ }\textbf {\bibinfo
  {volume} {35}},\ \bibinfo {pages} {1151--1160} (\bibinfo {year}
  {1998})}\BibitemShut {NoStop}%
\bibitem [{\citenamefont {Bucy}(2007)}]{Bucy2007}%
  \BibitemOpen
  \bibfield  {author} {\bibinfo {author} {\bibfnamefont {Brandon~R.}\
  \bibnamefont {Bucy}},\ }\emph {\bibinfo {title} {Student Understanding of
  Entropy and of Mixed Second-Order Partial Derivatives in Upper-Level
  Thermodynamics}},\ \href@noop {} {Ph.D. thesis},\ \bibinfo  {school}
  {University of Maine}, \bibinfo {address} {Orono, ME} (\bibinfo {year}
  {2007})\BibitemShut {NoStop}%
\bibitem [{\citenamefont {Langbeheim}\ \emph {et~al.}(2013)\citenamefont
  {Langbeheim}, \citenamefont {Safran}, \citenamefont {Livne},\ and\
  \citenamefont {Yerushalmi}}]{Langbeheim2013}%
  \BibitemOpen
  \bibfield  {author} {\bibinfo {author} {\bibfnamefont {Elon}\ \bibnamefont
  {Langbeheim}}, \bibinfo {author} {\bibfnamefont {Samuel~A.}\ \bibnamefont
  {Safran}}, \bibinfo {author} {\bibfnamefont {Shelly}\ \bibnamefont {Livne}},
  \ and\ \bibinfo {author} {\bibfnamefont {Edit}\ \bibnamefont {Yerushalmi}},\
  }\bibfield  {title} {\enquote {\bibinfo {title} {Evolution in students'
  understanding of thermal physics with increasing complexity},}\ }\href
  {\doibase 10.1103/PhysRevSTPER.9.020117} {\bibfield  {journal} {\bibinfo
  {journal} {Phys. Rev. ST Phys. Educ. Res.}\ }\textbf {\bibinfo {volume}
  {9}},\ \bibinfo {pages} {020117} (\bibinfo {year} {2013})}\BibitemShut
  {NoStop}%
\bibitem [{Note3()}]{Note3}%
  \BibitemOpen
  \bibinfo {note} {Data from the \protect \textit {Stat Mech}\ course at School
  1 were only analyzed for students who had previously taken \protect \textit
  {Thermo}.}\BibitemShut {Stop}%
\bibitem [{\citenamefont {Behrens}\ and\ \citenamefont
  {Smith}(2004)}]{Behrens2004}%
  \BibitemOpen
  \bibfield  {author} {\bibinfo {author} {\bibfnamefont {John~T.}\ \bibnamefont
  {Behrens}}\ and\ \bibinfo {author} {\bibfnamefont {Mary~Lee}\ \bibnamefont
  {Smith}},\ }\bibfield  {title} {\enquote {\bibinfo {title} {Data and data
  analysis},}\ }in\ \href@noop {} {\emph {\bibinfo {booktitle} {Handbook of
  Educational Psychology}}},\ \bibinfo {editor} {edited by\ \bibinfo {editor}
  {\bibfnamefont {David~C.}\ \bibnamefont {Berliner}}\ and\ \bibinfo {editor}
  {\bibfnamefont {Robert~C.}\ \bibnamefont {Calfee}}}\ (\bibinfo  {publisher}
  {Lawrence Erlbaum Associates, Inc},\ \bibinfo {address} {Mahwah, NJ},\
  \bibinfo {year} {2004})\ Chap.~\bibinfo {chapter} {30}, pp.\ \bibinfo {pages}
  {945--989}\BibitemShut {NoStop}%
\bibitem [{\citenamefont {Strauss}\ and\ \citenamefont
  {Corbin}(1990)}]{Strauss1990}%
  \BibitemOpen
  \bibfield  {author} {\bibinfo {author} {\bibfnamefont {A.}~\bibnamefont
  {Strauss}}\ and\ \bibinfo {author} {\bibfnamefont {J.}~\bibnamefont
  {Corbin}},\ }\href@noop {} {\emph {\bibinfo {title} {Basics of Qualitative
  Research: Grounded Theory Procedures and Techniques}}}\ (\bibinfo
  {publisher} {Sage},\ \bibinfo {address} {Thousand Oaks, CA},\ \bibinfo {year}
  {1990})\BibitemShut {NoStop}%
\bibitem [{\citenamefont {Heron}(2003)}]{Heron2003}%
  \BibitemOpen
  \bibfield  {author} {\bibinfo {author} {\bibfnamefont {Paula~R.L.}\
  \bibnamefont {Heron}},\ }\bibfield  {title} {\enquote {\bibinfo {title}
  {Empirical investigations of learning and teaching, part i: Examining and
  interpreting student thinking},}\ }in\ \href@noop {} {\emph {\bibinfo
  {booktitle} {Proceedings of the International School of Physics {Enrico
  Fermi,} Course CLVI: Research on Physics Education}}},\ \bibinfo {editor}
  {edited by\ \bibinfo {editor} {\bibfnamefont {Edward~F.}\ \bibnamefont
  {Redish}}\ and\ \bibinfo {editor} {\bibfnamefont {Matilde}\ \bibnamefont
  {Vicentini}}}\ (\bibinfo  {publisher} {IOS Press},\ \bibinfo {address}
  {Amsterdam},\ \bibinfo {year} {2003})\ pp.\ \bibinfo {pages}
  {341--350}\BibitemShut {NoStop}%
\bibitem [{\citenamefont {Smith}\ \emph {et~al.}(2009)\citenamefont {Smith},
  \citenamefont {Christensen},\ and\ \citenamefont {Thompson}}]{Smith2009}%
  \BibitemOpen
  \bibfield  {author} {\bibinfo {author} {\bibfnamefont {Trevor~I.}\
  \bibnamefont {Smith}}, \bibinfo {author} {\bibfnamefont {Warren~M.}\
  \bibnamefont {Christensen}}, \ and\ \bibinfo {author} {\bibfnamefont
  {John~R.}\ \bibnamefont {Thompson}},\ }\bibfield  {title} {\enquote {\bibinfo
  {title} {Addressing student difficulties with concepts related to entropy,
  heat engines, and the {C}arnot cycle},}\ }in\ \href@noop {} {\emph {\bibinfo
  {booktitle} {2009 Physics Education Research Conference, AIP Conference
  Proceedings}}},\ Vol.\ \bibinfo {volume} {1069},\ \bibinfo {editor} {edited
  by\ \bibinfo {editor} {\bibfnamefont {Mel}\ \bibnamefont {Sabella}}, \bibinfo
  {editor} {\bibfnamefont {Charles}\ \bibnamefont {Henderson}}, \ and\ \bibinfo
  {editor} {\bibfnamefont {Chandralekha}\ \bibnamefont {Singh}}},\ \bibinfo
  {organization} {American Association of Physics Teachers}\ (\bibinfo
  {publisher} {American Institute of Physics},\ \bibinfo {address} {Melville,
  NY},\ \bibinfo {year} {2009})\ pp.\ \bibinfo {pages} {277--281}\BibitemShut
  {NoStop}%
\bibitem [{\citenamefont {Smith}(2011)}]{Smith2011}%
  \BibitemOpen
  \bibfield  {author} {\bibinfo {author} {\bibfnamefont {Trevor~I.}\
  \bibnamefont {Smith}},\ }\emph {\bibinfo {title} {Identifying and Addressing
  Specific Student Difficulties in Advanced Thermal Physics}},\ \href
  {http://digitalcommons.library.umaine.edu/etd/263/} {Ph.D. thesis},\ \bibinfo
   {school} {University of Maine} (\bibinfo {year} {2011})\BibitemShut
  {NoStop}%
\bibitem [{\citenamefont {Erickson}(2006)}]{Erickson2006}%
  \BibitemOpen
  \bibfield  {author} {\bibinfo {author} {\bibfnamefont {Frederick}\
  \bibnamefont {Erickson}},\ }\bibfield  {title} {\enquote {\bibinfo {title}
  {Definition and analysis of data from videotape: Some research procedures and
  their rationales},}\ }in\ \href@noop {} {\emph {\bibinfo {booktitle}
  {Handbook of Complementary Methods in Education Research}}},\ \bibinfo
  {editor} {edited by\ \bibinfo {editor} {\bibfnamefont {J.~L.}\ \bibnamefont
  {Green}}, \bibinfo {editor} {\bibfnamefont {G.}~\bibnamefont {Camilli}}, \
  and\ \bibinfo {editor} {\bibfnamefont {P.~B.}\ \bibnamefont {Elmore}}}\
  (\bibinfo  {publisher} {Erlbaum},\ \bibinfo {address} {Mahwah, NJ},\ \bibinfo
  {year} {2006})\ Chap.~\bibinfo {chapter} {10}, pp.\ \bibinfo {pages}
  {177--205}\BibitemShut {NoStop}%
\bibitem [{Note4()}]{Note4}%
  \BibitemOpen
  \bibinfo {note} {In later versions students were asked about the working
  substance\ before being asked about the universe.}\BibitemShut {Stop}%
\bibitem [{\citenamefont {Everitt}\ and\ \citenamefont
  {Wykes}(1999)}]{Everitt1999}%
  \BibitemOpen
  \bibfield  {author} {\bibinfo {author} {\bibfnamefont {Brian~S.}\
  \bibnamefont {Everitt}}\ and\ \bibinfo {author} {\bibfnamefont {Til}\
  \bibnamefont {Wykes}},\ }\href@noop {} {\emph {\bibinfo {title} {A Dictionary
  of Statistics for Psychologists}}}\ (\bibinfo  {publisher} {Oxford University
  Press},\ \bibinfo {address} {New York, NY},\ \bibinfo {year} {1999})\ pp.\
  \bibinfo {pages} {69--70}\BibitemShut {NoStop}%
\bibitem [{\citenamefont {Denenberg}(1976)}]{Denenberg1976}%
  \BibitemOpen
  \bibfield  {author} {\bibinfo {author} {\bibfnamefont {Victor~H.}\
  \bibnamefont {Denenberg}},\ }\href@noop {} {\emph {\bibinfo {title}
  {Statistics and Experimental Design for Behavioral and Biological
  Researchers}}}\ (\bibinfo  {publisher} {John Wiley \& Sons, Inc.},\ \bibinfo
  {address} {New York, NY},\ \bibinfo {year} {1976})\ pp.\ \bibinfo {pages}
  {269--274}\BibitemShut {NoStop}%
\bibitem [{\citenamefont {Coladarci}\ \emph {et~al.}(2004)\citenamefont
  {Coladarci}, \citenamefont {Cobb}, \citenamefont {Minium},\ and\
  \citenamefont {Clarke}}]{Coladarci2004}%
  \BibitemOpen
  \bibfield  {author} {\bibinfo {author} {\bibfnamefont {Theodore}\
  \bibnamefont {Coladarci}}, \bibinfo {author} {\bibfnamefont {Casey~D.}\
  \bibnamefont {Cobb}}, \bibinfo {author} {\bibfnamefont {Edward~W.}\
  \bibnamefont {Minium}}, \ and\ \bibinfo {author} {\bibfnamefont {Robert~C.}\
  \bibnamefont {Clarke}},\ }\href@noop {} {\emph {\bibinfo {title}
  {Fundamentals of Statistical Reasoning in Education}}}\ (\bibinfo
  {publisher} {John Wiley \& Sons, Inc.},\ \bibinfo {year} {2004})\BibitemShut
  {NoStop}%
\bibitem [{Note5()}]{Note5}%
  \BibitemOpen
  \bibinfo {note} {When testing two populations for differences, the threshold
  for significance was set at $p<0.05$; when testing two populations for
  similarities, the threshold for significance was set at $p>0.10$. Values of
  $0.05<p<0.10$ are considered an indication of approaching significance (i.e.,
  not statistically significant but worth mentioning).}\BibitemShut {Stop}%
\bibitem [{Note6()}]{Note6}%
  \BibitemOpen
  \bibinfo {note} {The ``\protect \textit {Rev.\ + Irr.}'' and ``\protect
  \textit {Rev.\ + SF}'' categories are considered combinations of types of
  reasoning and not included in the ten primary types of reasoning. The
  ``\protect \textit {Statement}'' category is also not counted as a primary
  type of reasoning because it is evidenced by the absence of reasoning or
  explanation.}\BibitemShut {Stop}%
\bibitem [{\citenamefont {Baierlein}(1999)}]{Baierlein1999}%
  \BibitemOpen
  \bibfield  {author} {\bibinfo {author} {\bibfnamefont {Ralph}\ \bibnamefont
  {Baierlein}},\ }\href@noop {} {\emph {\bibinfo {title} {Thermal Physics}}}\
  (\bibinfo  {publisher} {Cambridge University Press},\ \bibinfo {address}
  {Cambridge, UK},\ \bibinfo {year} {1999})\BibitemShut {NoStop}%
\bibitem [{Note7()}]{Note7}%
  \BibitemOpen
  \bibinfo {note} {The efficiency of the Ralph engine changes with time because
  the reservoirs have finite heat capacities and changing temperatures.
  However, in the quasistatic approximation, the temperature of each reservoir
  does not change appreciably during any one cycle, so Ralph operates at the
  Carnot efficiency for the current temperatures. After many cycles, the
  temperatures of the reservoirs will be different, and CarnotÕs efficiency
  will also be different, but Ralph will always operate at CarnotÕs efficiency,
  and (more importantly) the combined entropy of the reservoirs remains
  constant.}\BibitemShut {Stop}%
\bibitem [{Note8()}]{Note8}%
  \BibitemOpen
  \bibinfo {note} {Some implementations of \protect \textit {Stat Mech}\ used
  an expanded version of Baierlein's problem 3.6 that included all parts of the
  FRQ shown in Fig.\ \ref {frq} \cite [p.\ 72]{Baierlein1999}.}\BibitemShut
  {Stop}%
\bibitem [{Note9()}]{Note9}%
  \BibitemOpen
  \bibinfo {note} {All student names are pseudonyms.}\BibitemShut {Stop}%
\bibitem [{Note10()}]{Note10}%
  \BibitemOpen
  \bibinfo {note} {For all transcripts, ``I'' stands for the
  Instructor.}\BibitemShut {Stop}%
\bibitem [{Note11()}]{Note11}%
  \BibitemOpen
  \bibinfo {note} {All transcriptions of differential vs.\ total change
  notation (d\relax $\@@underline {\hbox {\protect \hspace
  {0.5cm}}}\mathsurround \z@ $\relax vs.\ $\protect \mathrm {d}\protect \hspace
  {-0.5em}^-$\relax $\@@underline {\hbox {\protect \hspace
  {0.5cm}}}\mathsurround \z@ $\relax vs.\ $\Delta $\relax $\@@underline {\hbox
  {\protect \hspace {0.5cm}}}\mathsurround \z@ $\relax ) are directly from the
  video. For example, a transcription of ``$\protect \mathrm {d}\protect
  \hspace {-0.5em}^-Q$'' resulted from a verbal utterance of ``dee bar
  queue.''}\BibitemShut {NoStop}%
\bibitem [{\citenamefont {Stewart}(2001)}]{Stewart2001}%
  \BibitemOpen
  \bibfield  {author} {\bibinfo {author} {\bibfnamefont {James}\ \bibnamefont
  {Stewart}},\ }\href@noop {} {\emph {\bibinfo {title} {Calculus: Concepts and
  Contexts, Single Variable}}},\ \bibinfo {edition} {2nd}\ ed.\ (\bibinfo
  {publisher} {Brooks/Cole Publishing Co.},\ \bibinfo {year}
  {2001})\BibitemShut {NoStop}%
\bibitem [{\citenamefont {Orton}(1983)}]{Orton1983}%
  \BibitemOpen
  \bibfield  {author} {\bibinfo {author} {\bibfnamefont {A.}~\bibnamefont
  {Orton}},\ }\bibfield  {title} {\enquote {\bibinfo {title} {Students'
  understanding of differentiation},}\ }\href@noop {} {\bibfield  {journal}
  {\bibinfo  {journal} {Educational Studies in Mathematics}\ }\textbf {\bibinfo
  {volume} {14}},\ \bibinfo {pages} {235--250} (\bibinfo {year}
  {1983})}\BibitemShut {NoStop}%
\bibitem [{\citenamefont {Donaldson}(1963)}]{Donaldson1963}%
  \BibitemOpen
  \bibfield  {author} {\bibinfo {author} {\bibfnamefont {Margaret}\
  \bibnamefont {Donaldson}},\ }\href@noop {} {\emph {\bibinfo {title} {A Study
  of Children's Thinking}}}\ (\bibinfo  {publisher} {Tavistock Press},\
  \bibinfo {address} {London},\ \bibinfo {year} {1963})\BibitemShut {NoStop}%
\bibitem [{\citenamefont {Hu}\ and\ \citenamefont {Rebello}(2013)}]{Hu2013}%
  \BibitemOpen
  \bibfield  {author} {\bibinfo {author} {\bibfnamefont {Dehui}\ \bibnamefont
  {Hu}}\ and\ \bibinfo {author} {\bibfnamefont {N.~Sanjay}\ \bibnamefont
  {Rebello}},\ }\bibfield  {title} {\enquote {\bibinfo {title} {Understanding
  student use of differentials in physics integration problems},}\ }\href
  {\doibase 10.1103/PhysRevSTPER.9.020108} {\bibfield  {journal} {\bibinfo
  {journal} {Phys. Rev. ST Phys. Educ. Res.}\ }\textbf {\bibinfo {volume}
  {9}},\ \bibinfo {pages} {020108} (\bibinfo {year} {2013})}\BibitemShut
  {NoStop}%
\bibitem [{\citenamefont {Artigue}\ \emph {et~al.}(1990)\citenamefont
  {Artigue}, \citenamefont {Menigaux},\ and\ \citenamefont
  {Viennot}}]{Artigue1990}%
  \BibitemOpen
  \bibfield  {author} {\bibinfo {author} {\bibfnamefont {Mich\'{e}le}\
  \bibnamefont {Artigue}}, \bibinfo {author} {\bibfnamefont {Jacqueline}\
  \bibnamefont {Menigaux}}, \ and\ \bibinfo {author} {\bibfnamefont {Laurence}\
  \bibnamefont {Viennot}},\ }\bibfield  {title} {\enquote {\bibinfo {title}
  {Some aspects of students' conceptions and difficulties about
  differentials},}\ }\href@noop {} {\bibfield  {journal} {\bibinfo  {journal}
  {European Journal of Physics}\ }\textbf {\bibinfo {volume} {11}},\ \bibinfo
  {pages} {262--267} (\bibinfo {year} {1990})}\BibitemShut {NoStop}%
\end{thebibliography}%

\end{document}